\documentclass[aps, prd, superscriptaddress, nofootinbib, preprintnumbers, twocolumn]{revtex4-2}

\usepackage{uldm_pta}

\begin{document}

\title{Constraining Fundamental Constant Variations from Ultralight Dark Matter with Pulsar Timing Arrays}

\author{David E. Kaplan}
\email{david.kaplan@jhu.edu}
\affiliation{Department of Physics \& Astronomy, The Johns Hopkins University, Baltimore, MD, 21218, USA}
\author{Andrea Mitridate}
\email{amitri@caltech.edu}
\affiliation{Walter Burke Institute for Theoretical Physics, California Institute of Technology, Pasadena, CA 91125, USA}
\author{Tanner Trickle}
\email{ttrickle@caltech.edu}
\affiliation{Walter Burke Institute for Theoretical Physics, California Institute of Technology, Pasadena, CA 91125, USA}

\preprint{CALT-TH-2022-019}

\begin{abstract}

    Pulsar Timing Arrays (PTAs) are exceptionally sensitive detectors in the frequency band $\text{nHz} \lesssim f \lesssim \mu\text{Hz}$. Ultralight dark matter (ULDM), with mass in the range $10^{-23}\,\text{eV} \lesssim m_\phi \lesssim 10^{-20}\,\text{eV}$, is one class of DM models known to generate signals in this frequency window. While purely gravitational signatures of ULDM have been studied previously, in this work we consider two signals in PTAs which arise in presence of direct couplings between ULDM and ordinary matter. These couplings induce variations in fundamental constants, i.e., particle masses and couplings. These variations can alter the moment of inertia of pulsars, inducing pulsar spin fluctuations via conservation of angular momentum, or induce apparent timing residuals due to reference clock shifts. By using mock data mimicking current PTA datasets, we show that PTA experiments outperform torsion balance and atomic clock constraints for ULDM coupled to electrons, muons, or gluons. In the case of coupling to quarks or photons, we find that PTAs and atomic clocks set similar constraints. Additionally, we discuss how future PTAs can further improve these constraints, and detail the unique properties of these signals relative to the previously studied effects of ULDM on PTAs. 

\end{abstract}

\maketitle
\tableofcontents

\section{Introduction}
\label{sec:introduction}

Identifying the nature of dark matter (DM) is one of the most important goals of physics. While roughly five times more abundant than visible matter, shockingly little is known about how DM fits in the Standard Model (SM) of particle physics. One class of DM models, known as ultralight DM (ULDM), considers DM candidates with tiny particle masses, typically $m_\phi \ll 10^{-6} \, \text{eV}$. Such light DM candidates must be bosonic, otherwise they would severely violate the Tremaine-Gunn bound~\cite{Tremaine:1979we, DiPaolo:2017geq, Savchenko:2019qnn, Alvey:2020xsk} on light fermionic DM.

These light DM candidates are well described by classical fields oscillating in time with a frequency $2 \pi f \simeq m_\phi$. In the presence of a direct coupling between DM and the SM, these oscillations behave as time-dependent sources for SM fields, and can lead to time-dependent signals with $\omega \sim m_\phi$. Therefore Pulsar Timing Arrays (PTAs), thanks to their tremendous ability to detect signals in the $\text{nHz}-\mu \text{Hz}$ frequency range, offer a unique probe of ULDM candidates in the mass window $10^{-23}\, \text{eV}-10^{-20}\, \text{eV}$.\footnote{The lowest frequencies PTAs are sensitive to is set by the observation time, typically $\mathcal{O}(10 \, {\rm year})$, and the highest frequencies are set by the observation cadence, typically $\mathcal{O}({\rm week})$.} Models of ULDM in this mass range are theoretically interesting since they are able to suppress structure formation up to length scales much larger than a prototypical WIMP particle~\cite{Armengaud:2017nkf}. This feature of ULDM models has been proposed as a possible solution of the small-scale challenges faced by the $\Lambda$CDM paradigm, such as the missing satellites, and core-cusp problem. See Ref.~\cite{Bullock:2017xww} for a recent review.

The phenomenology of ULDM particles is vast~\cite{Arvanitaki:2014faa,Damour:2010rp,Hui:2016ltb,Damour:2010rm,Hu:2000ke,Marsh:2015xka,Arkani-Hamed:2002bjr, Centers:2019dyn}. Independent of any specific DM model, a cosmic background of ULDM will affect CMB observables~\cite{Hlozek:2014lca, Hlozek:2016lzm, Hlozek:2017zzf}, galaxy density profiles and rotation curves~\cite{Schive:2014dra, Schive:2014hza, Bar:2018acw, Bar:2021kti}, the Milky Way (sub)Halo mass functions~\cite{Nadler:2019zrb, Schutz:2020jox}, Lyman-$\alpha$ observables~\cite{Irsic:2017yje, Nori:2018pka, Armengaud:2017nkf}, and propagation of light due to metric fluctuations~\cite{Khmelnitsky:2013lxt, Kato:2019bqz, Porayko:2014rfa, Porayko:2018sfa}. Specific models of ULDM can introduce further interesting effects such as atomic clock shifts~\cite{Kennedy:2020bac,Hees:2016gop,Wcislo:2018ojh,VanTilburg:2015oza, Flambaum:2004tm,Arvanitaki:2014faa, Network2020}, relative accelerations in interferometers and PTAs~\cite{Graham:2015ifn, Morisaki:2020gui, Pierce:2018xmy, PPTA:2021uzb}, and changes to orbital periods of binary pulsars~\cite{Blas:2016ddr, Dror:2019uea, KumarPoddar:2019ceq}. Moreover, ULDM can mediate model dependent Yukawa forces between particles, allowing it to be searched for even in the absence of a cosmic background~\cite{Schlamminger:2007ht, Berge:2017ovy, Graham:2015ifn}.

CMB observables~\cite{Hlozek:2014lca, Hlozek:2016lzm, Hlozek:2017zzf} rule out ULDM with $m_\phi \lesssim 10^{-24} \, \text{eV}$ while recent studies of the Lyman-$\alpha$ forest~\cite{Irsic:2017yje, Nori:2018pka, Armengaud:2017nkf}, Milky Way (sub)Halo mass functions~\cite{Nadler:2019zrb, Schutz:2020jox}, and rotation curves~\cite{Bar:2018acw, Bar:2021kti} assert constraints of $10^{-21} \, \text{eV} \lesssim m_\phi$. However non-CMB constraints are susceptible to uncertainties related to determining properties of small scale structures, for which cosmological simulations~\cite{Zhang2019, Schive:2014hza}, and analytic methods~\cite{Press:1973iz} can be insufficient to derive bounds with high precision. Therefore it is important to have complementary probes in this mass region which are not subject to the same uncertainties.

In this work we focus on the effect of ULDM induced fluctuations of fundamental constants. These fluctuations, in the mass window considered here, have previously been searched for before via their effects on atomic clock systems~\cite{Kennedy:2020bac,Hees:2016gop,Wcislo:2018ojh,VanTilburg:2015oza, Network2020} and torsion balance experiments~\cite{Schlamminger:2007ht, Berge:2017ovy}. Here we detail two experimental signatures in PTAs first discussed in Ref.~\cite{Graham:2015ifn}. The first is due to fluctuations in the PTA reference clock. PTAs are exceptionally sensitive to timing residuals, and fluctuations in the reference clock used can be observed via these residuals. The second is from pulsar spin fluctuations. ULDM induced particle mass fluctuations will cause the pulsar mass to fluctuate. Therefore, by conservation of angular momentum, this must be accompanied by a fluctuation in the pulsar spin. We set constraints on these signals by using mock data closely resembling the current IPTA second data release, and a more futuristic PTA dataset. We find that in the mass range $10^{-23}\,\text{eV} \lesssim m_\phi \lesssim 10^{-20}\,\text{eV}$ current PTA constraints are competitive or stronger than current atomic clocks and fifth force constraints for a wide range of ULDM models.

The outline of this paper is as follows, in Sec.~\ref{sec:dm_signals} we begin by reviewing the ULDM models which give rise to fundamental constant fluctuations, and then show how these models generate a PTA signal by inducing pulsar spin fluctuations, Sec.~\ref{subsec:pulsar_spin_fluctuations}, and reference clock shifts, Sec.~\ref{subsec:reference_clock_shifts}. In Sec.~\ref{sec:analysis} we describe the analyses performed to set constraints, and the procedure followed to generate the mock data. In Sec.~\ref{sec:results} we discuss the results of the analyses, and the connection between the signals searched for here and those previously searched for in PTAs (Sec.~\ref{subsec:connection_to_similar_effects}).
 
\section{PTA signals from ULDM Induced Fundamental Constant Variations}
\label{sec:dm_signals}

As a model for ULDM we consider a singlet scalar field, $\phi$, which constitutes all the cosmic DM. Owing to their high phase space density, cold ($v_\phi \sim 10^{-3}$) ULDM candidates are well described by a classical field oscillating in time with a frequency $\omega = m_\phi$,
\begin{align}\label{eq:waveform}
	\phi(\vec{x}, t)= \frac{\sqrt{2 \rho_\phi}}{m_\phi} \hat{\phi} (\vec{x}) \cos{\left( m_\phi t + \gamma(\vec{x}) \right)} \, ,
\end{align}
where $\hat{\phi}(\vec{x})$ is a random amplitude with zero mean and unit variance, $\rho_\phi = 0.4\, \text{GeV}\,\text{cm}^{-3}$ is the local ULDM density, and $\gamma(\vec{x})$ is the phase of the field. 

This ULDM field can directly couple to ordinary matter in a plethora of ways. Following the notation in Ref.~\cite{Damour:2010rp}, we parameterize the couplings to the QED sector as
\begin{align}\label{eq:couplings}
	\mathcal{L}_{\phi, \rm QED}\supset\frac{\phi}{\Lambda}\left(\frac{ d_\gamma}{4e^2}F_{\mu\nu}F^{\mu\nu}-\sum_{f=e,\mu} d_{f} m_f\bar f f\right)\,,
\end{align}
where $\Lambda=M_{\rm Pl}/\sqrt{4\pi}$, $M_{\rm Pl}$ is the Planck mass, and $d$'s are dimensionless coupling coefficients.\footnote{Note that the electromagnetic field strength tensor, $F_{\mu\nu}$, is normalized such that the electron has unit charge.} Due to these couplings, fluctuations in the ULDM background induce variations in the fundamental constants of the SM. Specifically, the couplings in Eq.~\eqref{eq:couplings} drive fluctuations in the electromagnetic coupling constant, $\alpha$, and the electron and muon masses, $m_{e,\mu}$,
\begin{align}\label{eq:qed_parameter_shifts}
	\frac{\delta\alpha}{\alpha}=\frac{d_\gamma}{\Lambda} \phi \,,\qquad\frac{\delta m_{e,\mu}}{m_{e,\mu}}=\frac{d_{e, \mu}}{\Lambda} \phi\, .
\end{align}

ULDM can also couple to the QCD sector, however, more care must be taken relative to the QED case due to the running of masses and couplings (implicitly ignored for QED). Still following the notation of Ref.~\cite{Damour:2010rp}, we parameterize the couplings to the QCD sector as
\begin{align}\label{eq:qcd_couplings}
	\mathcal{L}_{\phi, \rm QCD}\supset\frac{\phi}{\Lambda}\left(\frac{d_g\beta_3}{2g_3}G_{\mu\nu}^AG_A^{\mu\nu}-\sum_{q=u,d}(d_q+\gamma_q d_g)m_q\bar qq\right)
\end{align}
where $\beta_3$ is the QCD beta function, and $\gamma_q$ are the light quark anomalous dimensions. This specific parameterization of the couplings is useful since it makes the $\phi$ induced shift to fundamental constants independent of $\beta_3$ and $\gamma_q$. Specifically, the light quark mass shifts are given by
\begin{align}
    \frac{\delta m_q}{m_q} & = \frac{d_q}{\Lambda} \phi\,,
    \label{eq:light_quark_mass_shift}
\end{align}
while the shifts of the nucleon masses are 
\begin{align}\label{eq:nucleon_shift}
	\frac{\delta m_{p, n}}{m_{p, n}}\simeq \frac{1}{\Lambda} \Big(d_g+C_n\,d_{\hat m}\Big) \phi \,,
\end{align}
where $C_{n}=0.048$~\cite{Damour:2010rp}, and we have defined the symmetric combination of the quark mass couplings as
\begin{align}
	d_{\hat m}\equiv\frac{d_{u} m_u+d_{d} m_d}{m_u+m_d}\,.
\end{align}
The first term in Eq.~\eqref{eq:nucleon_shift} comes from the shift to the QCD scale, $\delta \Lambda_\text{QCD} / \Lambda_\text{QCD} = d_g (\phi / \Lambda)$. The second, subleading, term comes from the shift to the light quark masses, Eq.~\eqref{eq:light_quark_mass_shift}. The contribution from a strange quark mass shift is expected to be smaller by a factor of $\sim4$~\cite{Flambaum:2004tm}, although there are uncertainties in deriving its precise value~\cite{Damour:2010rp}. The effect of the other $d$'s are even smaller \cite{Damour:2010rp}, and neglected here for simplicity, though in principle these searches can put constraints on those parameters as well.  

With the effects of ULDM on fundamental constants defined, we now describe how these can affect PTA observables. Pulsars are rotating, highly magnetized neutron stars that emit beams of electromagnetic radiation
from their magnetic poles. Given a misalignment between the rotation and magnetic axes, the pulsar rotation can cause the radiation beam to sweep across Earth. If this happens, a pulsar will appear to an Earth observer as a periodic emitter. Thanks to pulsars' extremely stable rotation periods, the time of arrival (TOA) of these radiation pulses can be predicted with great accuracy. PTAs accurately measure the TOAs by looking for deviations from these predictions, a quantity referred to as \textit{timing residuals}. In the following sections we will discuss how the fundamental constant variations can source these timing residuals and therefore be detected, or constrained, by PTAs.

Specifically, in Sec.~\ref{subsec:pulsar_spin_fluctuations} we describe how mass fluctuations can induce pulsars spin variations which naturally source timing residuals. In Sec.~\ref{subsec:reference_clock_shifts} we describe how shifts in fundamental constants can induce shifts in the energy levels of the atomic clocks used by PTAs, and lead to apparent aberrations in the pulsars TOAs. 

\subsection{Pulsar Spin Fluctuations}
\label{subsec:pulsar_spin_fluctuations}

We begin by studying pulsar spin fluctuations generated by changes in the pulsar moment of inertia, $I$. By conservation of angular momentum, these moment of inertia fluctuations will induce corresponding fluctuations in the pulsar spin frequency, $\omega$,
\begin{align}
    \frac{ \delta \omega}{\omega_0} =- \frac{ \delta I}{I_0}\,.
\end{align}
Fluctuations in the pulsar constituent particle (dominantly neutrons, protons, electrons, and muons) masses can induce fluctuations in $I$. This dependence comes implicitly via the pulsar mass, $M$, as well as explicitly via the neutron mass which controls the balance of the Fermi degeneracy and gravitational pressures. In the simplest model of a spherically symmetric, non-rotating neutron star consisting of only non-relativistic neutrons one can show that $R \propto M^{-1/3} m_n^{-8/3}$, and therefore, $I \propto M^{1/3} m_n^{-16/3}$. This is obviously a simplistic description, and a more realistic model for the mass dependence of $I$ would require numerically solving the Tolman-Oppenheimer-Volkov equations~\cite{Lim:2018xne}, and include relativistic corrections. These corrections will induce $\mathcal{O}(1)$ deviations from the naive scaling; because of this we introduce the $\eta$ and $\delta \eta$, pulsar dependent, parameters such that 
\begin{align}
    \frac{ \delta I}{I_0} = \eta \frac{\delta M}{M_0} + \delta \eta \frac{\delta m_n}{m_n}\,.
\end{align}
In the following, for simplicity, we will assume that $\eta = 1/3$, $\delta \eta = -16/3$, the values computed from the simplest model.

These fluctuations in the pulsar spin frequency induce corresponding fluctuations, $h$, in the pulsar timing residuals via,
\begin{align}\label{eq:df2h}
	h= -\int\frac{\delta\omega}{\omega_0}dt=\int\frac{\delta I}{I_0}dt\,.
\end{align}

From here it is clear how to understand the effects of ULDM models on pulsar timing; one just has to relate the particle mass fluctuations, $\delta m_f/m_f$, to pulsar mass fluctuations, $\delta M / M_0$,
\begin{align}\label{eq:p_mass_shift}
    \frac{\delta M}{M_0} & = \sum_{f \in \{ e, \mu, p, n \} } Y_f \frac{m_f}{m_n} \frac{\delta m_f}{m_f}\,,
\end{align}
where $Y_f \equiv N_f/(N_n + N_p)$ is the relative number of $f$ particles to protons and neutrons in the pulsar. While these values will vary from pulsar to pulsar, we take $Y_n \sim 0.9, Y_p \sim 0.1, Y_e = Y_p$ and $Y_\mu \sim 0.05$ as fiducial values following Ref.~\cite{Bell:2019pyc}. 

We can now substitute Eqs.~\eqref{eq:qed_parameter_shifts},~\eqref{eq:nucleon_shift} in to Eq. \eqref{eq:p_mass_shift} to finally obtain the ULDM induced timing residuals in terms of the DM background field:
\begin{align}
    h(t) = & \frac{\sqrt{2 \rho_\phi}}{m_\phi^2 \Lambda} \left( \vec{y} \cdot \vec{d} \,\right) \hat{\phi}_P \sin{\big( m_\phi t + \gamma(\vec{x}_P) \big)}\,,
        \label{eq:pulsar_spin_fluctuation_signal}
\end{align}
where $\vec{x}_P$ is the position of the pulsar, and $\vec{y} \cdot \vec{d} \equiv \sum_{i} y_i d_i$ where $i \in \{ e, \mu, \gamma, g, \hat{m} \}$ and we have introduced the sensitivity parameters of this search:
\begin{equation}
    \label{eq:y_PTA_PSF}
    \begin{split}
        \{y_g,\,y_{\hat m}, \,y_\mu,\,y_e\}=&\,\eta\left\{1,\,C_n,\, 6\times10^{-3},\, 5\times 10^{-5}\right\} \\ 
        +&\delta \eta\left\{1,\,C_n,\,0,\,0\right\} \, .
    \end{split}
\end{equation}
One noteworthy feature is that while muons are less abundant than electrons, due to their larger mass the sensitivity is greater.

\subsection{Reference Clock Shifts}
\label{subsec:reference_clock_shifts}

The TOAs measured by PTAs are referenced to the Temps Atomique International (TAI) realization of the Terrestrial Time (TT). The TAI estimates TT using measurements from an ensemble of more than 400 atomic clocks (most of them based on the hyperfine transition of the ground state of the Cesium atom). A shift in the frequency of these clocks will induce an apparent shift in the pulsar spin frequency, which translates to pulsar timing residuals according to the first equality in Eq.~\eqref{eq:df2h}. In this section, we will discuss how fundamental constant variations can induce such shifts.

Using a similar notation to Ref.~\cite{Flambaum:2004tm}, the scaling of the frequency of an atomic clock is given by 
\begin{align}
    f \propto \Big(m_e \alpha^2 \Big)\Big[\alpha^2 F_{\rm rel}(Z\alpha)\Big]\left(\mu \frac{m_e}{m_p}\right)^\zeta
    \label{eq:clock_scaling}
\end{align}
where $F_{\rm rel}(Z\alpha)$ is the relativistic correction to the energy levels of an atom with nuclear charge $Z$, $\mu$ is the nuclear magnetic moment, and $\zeta=1$ ($\zeta=0$) for clocks using hyperfine (optical) transitions. It is clear then that fluctuations in fundamental constants will induce fluctuations in atomic clock frequencies. Specifically, for a clock using an atom $A$, 
\begin{equation}\label{eq:clock_f_shift}
\begin{split}
	\frac{\delta f_A}{f_A}\simeq&\Bigg[\frac{\delta m_e}{m_e}+(4+K_A)\frac{\delta\alpha}{\alpha}\\
	 &+\zeta\Bigg(\frac{\delta m_e}{m_e}+C_A\sum_{q=u,d}\frac{\delta m_q}{m_q}-\frac{\delta m_p}{m_p}\Bigg)\Bigg]\, ,
\end{split}
\end{equation}
where $\delta F_{\rm rel}/F_{\rm rel}=K_A\,\delta\alpha/\alpha$, and $\delta \mu/\mu=C_A\,\delta m_q/m_q$. For the case of Cesium atoms, $K_A=0.83$, and $C_A=0.110$ \cite{Flambaum:2004tm}. 

Finally, by substituting Eqs.~\eqref{eq:qed_parameter_shifts},~\eqref{eq:nucleon_shift} in to Eq.~\eqref{eq:clock_f_shift} we obtain the induced timing residuals in terms of the ULDM background:
\begin{align}
    h(t) = \frac{\sqrt{2 \rho_\phi}}{m_\phi^2 \Lambda} \left( \vec{y} \cdot \vec{d} \right) \hat{\phi}_E \sin{\big( m_\phi t + \gamma(\vec{x}_E) \big)} \, .
    \label{eq:reference_clock_shift_signal}
\end{align}
where $\vec{x}_E$ is the position of the Earth, and the sensitivity parameters of this search are given by:
\begin{align}
    \label{eq:y_PTA_clock}
	\left\{y_g,\,y_\gamma,\,y_{\hat m},\,y_e\right\}\simeq\left\{\zeta,\,\xi_A,\,\zeta\left( C_n + \hat C_A \right),1+\zeta\right\}\,,
\end{align}
where $\xi_A\equiv4+K_A$, and $\hat C_A=C_A (m_u+m_d)^2/ 2m_u m_d$.\\ 

While the signals in Eq.~\eqref{eq:pulsar_spin_fluctuation_signal} and Eq.~\eqref{eq:reference_clock_shift_signal} appear similar, they have a few distinguishing features. First, notice the difference in the sensitivity parameters, $\vec{y}$ in Eqs.~\eqref{eq:y_PTA_clock},~\eqref{eq:y_PTA_PSF}. The pulsar spin fluctuation signal is dominant for muon-philic models, while the reference clock shift is dominant for models coupling to photons or electrons. The two signals also differ in the correlations between pulsars. Assuming the distance between pulsars is much larger than the DM coherence length, the pulsar spin fluctuations will be uncorrelated between pulsars, i.e. $\langle h_I h_J \rangle \propto \langle \hat{\phi}(\vec{x}_I) \hat{\phi}(\vec{x}_J) \rangle \simeq \delta_{IJ}$. Conversely, the reference clock shift affects all of the pulsars in an identical way; this will leave an imprint of monopole correlations across the array, or $\langle h_I h_J \rangle \propto 1$. Note that both of these signals are distinct from those of gravitational waves or Doppler shifts which will leave quadrupole, and dipole signatures, respectively. A summary of the difference between these effects and other ULDM effects in PTAs is given in Table~\ref{tab:effect_summary} in Sec.~\ref{subsec:connection_to_similar_effects}. 

\begin{table*}[t]
\begin{center}
\renewcommand{\arraystretch}{1.3}
\setlength{\tabcolsep}{6.5pt}
\small
\begin{tabular}{llll}
\toprule
\multicolumn{1}{c}{\bf{Parameter}}  & \multicolumn{1}{c}{\bf{Description}} & \multicolumn{1}{c}{\bf{Prior}} & \multicolumn{1}{c}{\bf{Comments}} \\
\midrule

\multicolumn{4}{c}{\bf{White Noise}} \\[1pt]
$E_{\mu}$ & EFAC per backend/receiver system & Uniform $[0.1, 5]$ & one parameter per pulsar \\
$Q_{\mu}$ [s] & EQUAD per backend/receiver system & log-Uniform $[-8.5, -5]$ & one parameter per pulsar  \\
\midrule

\multicolumn{4}{c}{\bf{Red Noise}} \\[1pt]
$A_{\rm red}$ & red noise power-law amplitude & log-Uniform $[-20, -11]$ & one parameter per pulsar  \\
$\gamma_{\rm red}$ & red noise power-law spectral index & Uniform $[0, 7]$ & one parameter per pulsar \\
\midrule

\multicolumn{4}{c}{\bf{ULDM}} \\[1pt]
$A_{i}$ & ULDM signal amplitude & log-Uniform $[-20, -14]$ & one parameter for PTA \\
$m_\phi\,[{\rm eV}]$ & ULDM mass & log-Uniform $[-24, -19]$ & one parameter for PTA \\
$\hat\phi_{E}^2$ & Earth normalized signal amplitude & $e^{-x}$ & one parameter per PTA \\
$\hat\phi_{P}^2$ & pulsar normalized signal amplitude & $e^{-x}$ & one parameter per pulsar$^*$ \\
$\gamma_{E}$ & Earth signal phase & Uniform $[0, 2\pi]$ & one parameter per PTA \\
$\gamma_{P}$ & pulsar signal phase & Uniform $[0, 2\pi]$ & one parameter for pulsar\\
\midrule

\multicolumn{4}{c}{\bf{Supermassive Black Bole Binaries (SMBHB)}} \\[1pt]
$A_{\mathrm{GWB}}$ & common process strain amplitude & log-Uniform $[-18, -14]$ & one parameter for PTA \\
$\gamma_{\mathrm{GWB}}$ & common process power-law spectral index & delta function ($\gamma_\mathrm{GWB}=13/3$)& fixed \\

\bottomrule

\end{tabular}
\caption{Prior distributions for the parameters used in all the analyses in this work. The $^*$ indicates parameters that are present only in the uncorrelated analyses. \label{tab:priors}}
\end{center}
\end{table*}
\section{Data Analysis}
\label{sec:analysis}

In this section, we discuss the data analysis techniques used to constrain ULDM couplings. The analysis will closely follow the standard procedure adopted by PTA collaborations. This section aims to summarize how this procedure can be applied in our context. See Ref.~\cite{NANOGrav:2015aud} for a more detailed discussion on Bayesian inference with PTA data, as well as Ref.~\cite{Taylor:2021yjx} for a more pedagogical discussion.

\subsection{Noise Modeling and the PTA Likelihood}

The main observable in a PTA experiment is the timing residuals, $\vec{\delta t}$, which measure the discrepancy between the observed times of arrival (TOAs) and the ones predicted by the pulsar timing model every cadence, $\Delta t$. Generally, there are three main contributions to these timing residuals: white noise, red noise, and small errors in the fit to the timing-ephemeris parameters. Specifically, we can model the timing residuals as:
\begin{align}
	\vec{\delta t}=\vec{n}+\mat{\rm F}\vec{a}+\mat{\rm M}\vec{\epsilon}\,.
\end{align}
We will now discuss each of these three terms in more detail.

For each of the $N$ TOAs, the white noise is assumed to be a normally distributed random variable, with zero mean and variance,
\begin{align} 
	\langle n_{i,\mu} n_{j,\nu}\rangle	=E_\mu^2\sigma_i\delta_{ij}\delta_{\mu\nu}+Q_\mu^2\delta_{ij}\delta_{\mu\nu}
\end{align}
where $\sigma_i$ is the TOA uncertainty for the $i-$th observation, $E_\mu$ is the {\it Extra FACtor} (\texttt{EFAC}) parameter for the receiver-backend system $\mu$, and $Q_\mu$ is the {\it Extra QUADreature} (\texttt{EQUAD}) parameter.\footnote{If the data are measured in multiple frequency bands, as they are in NANOGrav, there is also {\it Extra CORRelated} (\texttt{ECORR}) white noise between different frequency bands within the same observation epoch \cite{NANOGrav:2014zwv}. It should also be noticed that when considering data sets with much longer observation times, and using telescopes with greater sensitivity than current PTAs, \texttt{EFAC} and \texttt{EQUAD} might not capture the effects from all noise sources \cite{Cordes:2013iea}.}

Red noise is modeled in a Fourier basis for frequencies $k/T$, where $k$ indexes the harmonics of the basis and $T$ is the observation time. Since red noise is a low-frequency process, the summation over $k$ is truncated to some reasonable frequency, $N_f$. For example, in the most recent NANOGrav search for a gravitational wave background (GWB) this cutoff was set to $N_f=30$. We will use the same cutoff in our analyses. This set of $2N_f$ sine-cosine pairs evaluated at the different observation times is contained in the {\it Fourier density matrix}, $\mat{\rm F}$. The Fourier coefficients $\vec{a}$ are assumed to be normally distributed random variables, with zero mean and covariance matrix $\langle\vec{a}\,\vec{a}^T\rangle=\mat{\phi}$. This covariance matrix will contain all the possible sources of low-frequency achromatic noise. For our analysis we will consider two possible sources: pulsar intrinsic red noise, and GWB. Therefore, $\mat{\phi}$ takes the form 
\begin{align}
	[\phi]_{(ak)(bj)}=\Gamma_{ab}\rho_k\delta_{kj}+\kappa_{ak}\delta_{kj}\delta_{ab}
\end{align}
where $a$ and $b$ index pulsars, $k$ and $j$ index frequency harmonics, and $\Gamma_{ab}$ is the GWB overlap function for the pulsar pair $(a,b)$. In this expression the terms $\rho_k$ ($\kappa_{ak}$) are related to the power spectral density (PSD) of the timing residuals, $S(f)$, induced by the GWB (pulsar intrinsic red noise) such that $\rho(f)=S(f)\Delta f$ with $\Delta f=1/T$ (and similarly for $\kappa_a(f)$). Following the assumptions commonly made in PTA analysis, we model the PSD for the two red noise contributions as 
\begin{align}
	\rho(f)&=\frac{A_{\rm GWB}^2}{12\pi^2}\left(\frac{f}{1 \, {\rm year}^{-1}}\right)^{-\gamma_{\rm GWB}}\,{\rm year}^2\,,\\
	\kappa(f)&=\frac{A_a^2}{12\pi^2}\left(\frac{f}{1 \, {\rm year}^{-1}}\right)^{-\gamma_a}\,{\rm year}^2\,.
\end{align}

Finally, possible small deviations from the initial best-fit values of the $m$ timing-ephemeris parameters are accounted for by the term $\mat{M}\vec{\epsilon}$. The {\it design matrix}, $\mat{M}$, is an $N\times m$ matrix containing the partial derivatives of the TOAs with respect to each timing-ephemeris parameter (evaluated at the initial best-fit value), and $\vec{\epsilon}$ is a vector containing the linear offset from these best-fit parameters. 

Since in most analyses, ours included, we do not care about the specific realization of the noise but are instead interested in its statistical properties, we can analytically marginalize over all the possible noise realizations (i.e., integrate over all the possible values of $\vec{a}$ and $\vec{\epsilon}$). This leaves us with a marginalized likelihood that depends only on the hyper-parameters, $\eta=(A_{\rm GWB}, \gamma,\ldots)$, characterizing the statistical properties of the noise \cite{Lentati:2012xb, vanHaasteren:2012hj}: 
\begin{align}\label{eq:likelihood}
	p(\vec{\delta t}|\eta)=\frac{\exp\left(-\frac{1}{2}\vec{\delta t}^T\mat{C}^{-1}\vec{\delta t}\right)}{\sqrt{{\rm det}(2\pi\mat{C})}}\,,
\end{align}
where $\mat{C}=\mat{N}+\mat{TBT}^T$. Here $\mat{N}$ is the covariance matrix of the white noise, $\mat{B}={\rm diag}(\mat{\infty},\mat{\phi})$, and $\mat{T}=[\mat{M},\mat{F}]$.

\subsection{Signal Parameterization}
The marginalized likelihood in Eq.~\eqref{eq:likelihood} can be easily generalized to take into account deterministic signals in the timing residuals. In presence of a deterministic signal, $\vec{h}(\vec{\theta})$, which depends on the parameters $\vec{\theta}$, we just need to make the replacement $\vec{\delta t}\to\vec{\delta t}-\vec{h}(\vec{\theta})$ in Eq.~\eqref{eq:likelihood}. In our case the deterministic signal is given by the ULDM induced fluctuations described in Sec.~\ref{sec:dm_signals}. 

Generally, the ULDM signals for the $I^\text{th}$ pulsar in the array will have the form
\begin{widetext}
\begin{equation}\label{eq:h_parametrization}
	h_I(t)=\frac{A_i}{m_\phi}\bigg[y_E^i\,\hat\phi_E\sin\big(m_\phi t+\gamma_E\big)+y_P^i\,\hat\phi_{P,I}\sin\big(m_\phi t+\gamma_{P,I}\big)\bigg]
\end{equation}
\end{widetext}
where $A_i=d_i\sqrt{2 \rho_\phi}/(m_\phi \Lambda)$ here, and we have defined the Earth and pulsars phases as 
\begin{align}
    \gamma_E&=\vec{k}\cdot\vec{x}_E+\tilde\gamma_E\,,\\
    \gamma_{P,I}&=\vec{k}\cdot\vec{x}_{P,I}-m_\phi |\vec{x}_{P,I}-\vec{x}_E|+\tilde\gamma_{P,I}\,.
\end{align}
where $k\sim m_\phi v_\phi$ is the characteristic DM momentum. 

We will now briefly discuss how we treat the parameters appearing in Eq.~\eqref{eq:h_parametrization}. Since the observation times of interest here are $T\sim10\,{\rm year}$, the motion of celestial bodies is of order $\mathcal{O}(10^{-3}\,{\rm pc})$ and we can safely ignore the variation in $\vec{k}\cdot\vec{x}$ due to the Earth/pulsar motion and take $\gamma_{E,P}$ to be constants. Moreover, since typical values of $|\vec{x}_P-\vec{x}_E|$ range between $0.1$ and several kpc, the term $m_\phi|\vec{x}_E-\vec{x}_P|$ is never negligible for the DM masses of interest here. Therefore, $\gamma_P$ and $\gamma_E$ need to be treated as independent parameters. The DM momentum sets the coherence length, $l_c$, of the DM background,
\begin{align}
	l_c\simeq\frac{2\pi}{m_\phi v_\phi}\sim0.4\,{\rm kpc}\left(\frac{10^{-22}\,{\rm eV}}{m_\phi}\right)\,.
\end{align}
The DM background amplitude, $\hat{\phi}(\vec{x})$, is correlated within a coherence length, and uncorrelated outside of it. Since the typical pulsar-pulsar and pulsar-Earth separations are comparable with the ULDM coherence length, following the same practice of \cite{PPTA:2021uzb}, we perform our analysis in two limits:
\begin{itemize}
    \item {\it Correlated}: in this case we will have one $\hat\phi$ parameter for both the Earth and all the pulsars terms: $\hat\phi_E=\hat\phi_{P,I}$ for all $I$
    \item {\it Uncorrelated}: in this case $\hat\phi_E$ and all the $\hat\phi_{P,I}$ will be uncorrelated free parameters
\end{itemize}
The {\it correlated} analysis is expected to provide reliable results for the low-mass region considered in this work, and vice-versa for the {\it uncorrelated} analysis. 

\subsection{Setting Constraints}
 
To set constraints, we derive the Bayesian posteriors distributions for the noise and DM parameters by using the Markov Chain Monte Carlo (MCMC) techniques implemented in the \texttt{PTMCMCSampler} package \cite{2019ascl.soft12017E}, together with the marginalized likelihood given in Eq.~\eqref{eq:likelihood} (implemented by using the \texttt{enterprise} \cite{2019ascl.soft12015E} and \texttt{enterprise\_extensions} \cite{enterprise} packages), and the priors distributions given in Table \ref{tab:priors}. We then marginalize over all parameters except the DM signal amplitude, $A_i$, and the DM mass, $m_\phi$. The constraints on the signal amplitude in each mass bin are then set to the 95$^\text{th}$ percentile of the amplitude in that bin. These constraints can then be easily translated to constraints on the DM couplings $d_i$.

\subsection{Mock Data}

While future analyses will be performed using NANOGrav and IPTA data, in this work we will set constraints by using mock data that should closely resemble the current IPTA DR2 dataset \cite{Perera:2019sca} and future PTAs data \cite{https://doi.org/10.48550/arxiv.1501.00056, https://doi.org/10.48550/arxiv.1907.07648, Hobbs:2014tqa}. The mock data are generated by using the pulsar parameters contained in the \texttt{.par} files of the International Pulsar Timing Array (IPTA) second Mock Data Challenge (MDC2) \cite{Hazboun:2018wpv}, the python wrapper \texttt{libstempo} \cite{2020ascl.soft02017V} and the pulsar timing package \texttt{TEMPO2} \cite{Hobbs:2006cd, Edwards:2006zg}.
 
For each pulsar we generate a list of TOAs by using the PTA parameters ({\it i.e.} cadence, $\Delta t$, TOAs uncertainties, $\sigma_i$, number of pulsars, $N_P$, and observation time, $T$) given in Table \ref{tab:pta_parameters}. The TOAs are injected with the noise sources described in the previous sections: white noise, pulsar intrinsic red noise, and GWB. For each pulsar, the values of the noise parameters are randomly sampled from the priors given in Table \ref{tab:priors}.
\begin{table}[t]
\bgroup
\renewcommand{\arraystretch}{1.5}
\setlength\tabcolsep{8pt}
	\begin{tabular}{ccccc}
		\toprule
					& 	$N_P$ \qquad	& 	$\sigma\,[{\rm ns}]$	&	$T$ [year]  	  	& 	$\Delta t$ [week]     \\ \hline
		IPTA			&	$65$			&	100					&	$15$ 		&	$3$	 \Tstrut\Bstrut\\ 
		Future PTA			&	$200$		&	50					&	$20$			&	$2$	\Tstrut\Bstrut \\
        \bottomrule
	\end{tabular}
\egroup
\caption{PTAs parameters used to generate mock data. Specifically: number of pulsars, $N_P$, TOA uncertainties, $\sigma_i$, observation time, $T$, and observation cadence, $\Delta t$.\label{tab:pta_parameters} }
\end{table} 

\section{Results}
\label{sec:results}

\begin{figure*}
    \includegraphics[width=\textwidth]{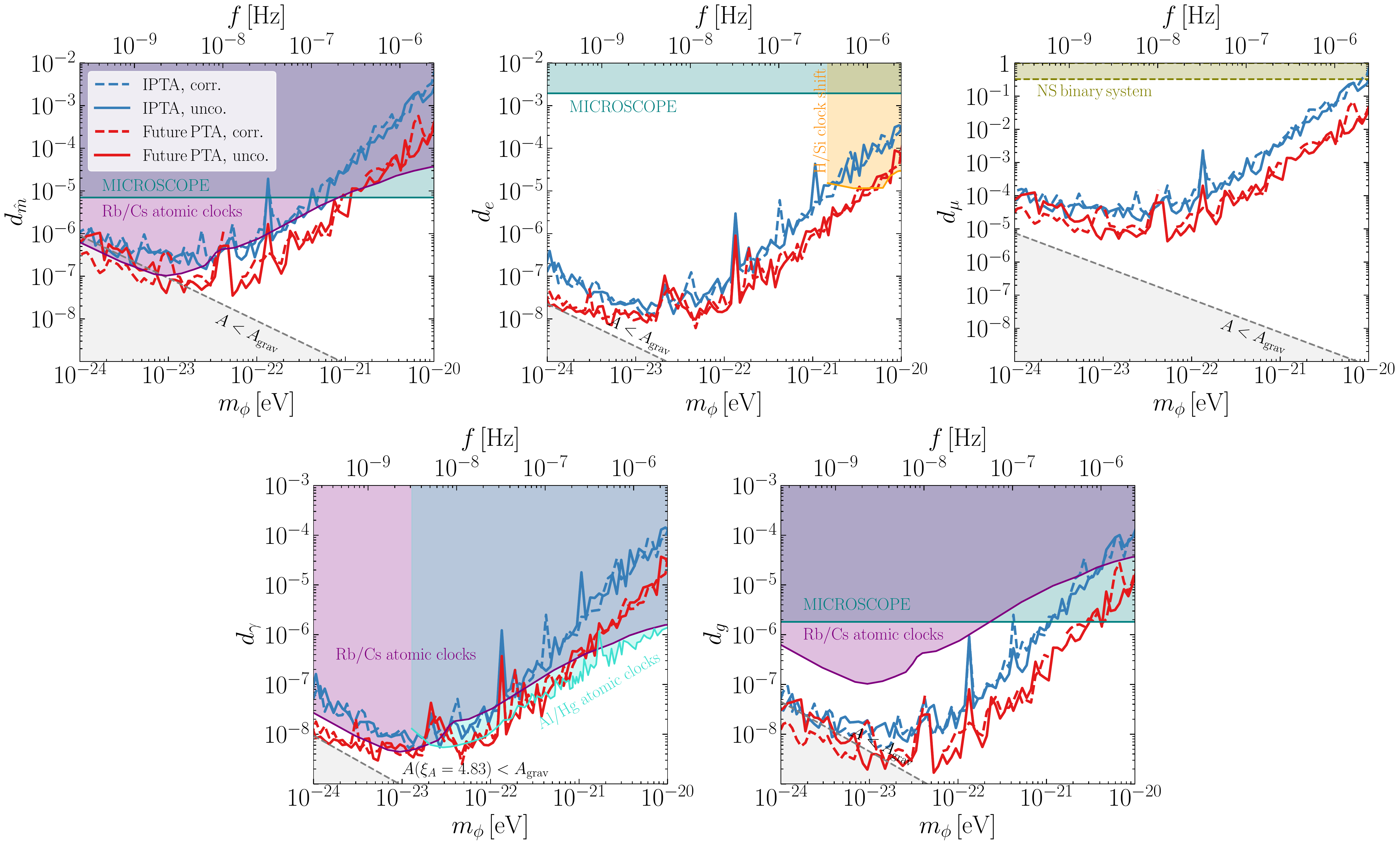}
    \caption{The blue (red) lines show the 95\% C.L. constraints on five models of ULDM derived by using a mock IPTA (future PTA) dataset with PTA parameters given in Table~\ref{tab:pta_parameters}. Dashed (solid) lines correspond to searches for an (un)correlated signal (labelled in ``unco.", ``corr.", respectively), as discussed in Sec.~\ref{sec:analysis}. In deriving these plots we have set $\eta=1/3$, $\delta \eta = -16/3$, and assumed that PTA clocks use the transition between the two hyperfine levels of the $^{133}{\rm Cs}$ ground state. The lower gray shaded regions correspond to regions of parameter space where the signal amplitude is less than the purely gravitational signal. Current constraints ``Rb/Cs atomic clocks" (purple) are from Ref.~\cite{Hees:2016gop}, ``Al/Hg atomic clocks" (turquoise) are from Ref.~\cite{Network2020}, ``MICROSCOPE" (teal) are from Ref.~\cite{Berge:2017ovy}, ``H/Si clock shift" (orange) are from Ref.~\cite{Kennedy:2020bac}, and ``NS binary system" are from Refs.~\cite{Dror:2019uea, KumarPoddar:2019ceq}.}
    \label{fig:constraints}
\end{figure*}

Applying the analysis detailed in Sec.~\ref{sec:analysis} to the signals discussed in Sec.~\ref{sec:dm_signals}, we place projected constraints on the five scalar ULDM coupling constants: $d_{\hat{m}}, d_e, d_\mu, d_\gamma$ and  $d_g$. The results are shown in Figs.~\ref{fig:constraints}, where we compare them to constraints from atomic clocks~\cite{Hees:2016gop, Kennedy:2020bac, Network2020}, torsion balances \cite{Berge:2017ovy}, and decay of neutron star binaries orbits \cite{Dror:2019uea, KumarPoddar:2019ceq}. While other constraints apply to these models (see, e.g., Refs.~\cite{Janish:2020knz, Berge:2017ovy, Kennedy:2020bac}) to simplify Fig.~\ref{fig:constraints} we only show those which have the strongest constraints in the mass range of interest here: $10^{-24}\, \text{eV} \lesssim m_\phi \lesssim 10^{-20} \, \text{eV}$. 

We find that, in the mass range under consideration, an IPTA-like array can already be competitive with the most precise atomic clock experiments~\cite{Hees:2016gop, Kennedy:2020bac, Network2020} available today when ULDM couples to the QCD sector via $d_{\hat{m}}$, and strictly outperforms them when the coupling happens through $d_g$. In the QED sector, atomic clock experiments are slightly better at constraining $d_\gamma$.\footnote{The constraints derived in Ref.~\cite{Network2020} (as well as Ref.~\cite{PPTA:2021uzb}) take into account the stochastic nature of $\hat{\phi}$, as done here. Some previous constraints, e.g., those in Ref.~\cite{Hees:2016gop}, assume $\hat{\phi} = 1$ which can lead to $\mathcal{O}(1)$ differences in the projections~\cite{Centers:2019dyn}. For simplicity, we directly reproduce the constraints from their respective paper and do not rescale them.} However, they are typically not able to constrain $d_e$ since relative frequency shifts between clocks, using the same type of atomic transition, cause the $d_e$ dependence to drop out. To avoid this cancellation one needs to compare different types of transitions as done in~\cite{Kennedy:2020bac}. At the moment, the observation time of this experiment limits the constraints to $10^{-21}\,{\rm eV} \lesssim m_\phi$. PTAs can also beat torsion balance constraints~\cite{Schlamminger:2007ht, Berge:2017ovy} by several orders of magnitude. Lastly, we note that PTAs are exceptionally sensitive to $d_\mu$. This is due to the large number of muons within neutron stars, sustained through $\beta$ equilibrium~\cite{Garani:2019fpa, Zhang:2020wov}. This allows PTAs to improve by more than three orders of magnitude the next best projected constraints from the orbital decay of a neutron star binary system~\cite{Dror:2019uea, KumarPoddar:2019ceq}.

All of the previous conclusions are amplified for future PTA experiments. With better timing parameters and more pulsars, a future PTA will be a strict improvement over the IPTA, as can be seen in Fig.~\ref{fig:constraints}.\footnote{The increased dimensionality of the parameter space for the future PTA setup makes sampling computationally demanding. Because of this, some constraints are less converged, specifically the correlated constraints on $d_{\hat m},\,d_\mu$, and $d_g$. In Fig.~\ref{fig:constraints}, these constraints have been smoothed by averaging the constraints from neighboring mass points.} Similar improvements could also be achieved by the IPTA collaborations, as the observation time and number of pulsars keep increasing. To gain intuition for the dependence of these constraints on the PTA parameters we will momentarily resort to a frequentist signal analysis approach, see Refs.~\cite{Ramani:2020hdo,Moore:2014lga,Smith:2019wny,Allen:1997ad} for more details. Given a deterministic signal in each pulsar, $h_I = A \sin{(m_\phi t + \gamma)}$ for simplicity, and assuming $N_P$ identical pulsars with only white noise, the signal SNR can be shown to scale as,
\begin{align}
    \text{SNR}^2 & \propto \frac{N_P T}{\sigma^2 \Delta t} A^2 \, .
    \label{eq:snr}
\end{align}
This explains the scaling of the constraints in Fig.~\ref{fig:constraints} at high masses; $\text{SNR} \propto A \propto d / m_\phi^2$ and therefore the constraints on $d$ are proportional to $m_\phi^2$. The scaling of Eq.~\eqref{eq:snr} is appropriate when the signal is known in all the pulsars, i.e., for an Earth term signal, since this SNR is derived using a matched filter approach. A pulsar dependent signal would have a more complicated, weaker, dependence on $N_P$, with all other dependencies being equal. With the scaling in Eq.~\eqref{eq:snr} future PTAs are expected to perform better than IPTA-like ones by a factor,
\begin{align}
    \frac{d}{d_\text{IPTA}} \sim \left( \frac{65}{N_P} \right)^\frac{1}{2} \left( \frac{15 \, \text{year}}{T} \right)^\frac{1}{2} \left( \frac{\sigma}{100 \, \text{ns}} \right) \left( \frac{\Delta t}{3 \, \text{week}} \right)^\frac{1}{2} \, .
    \label{eq:pta_scaling}
\end{align}
For the specific future PTA this means a predicted improvement of a factor of $\sim 5$, in good agreement with Fig.~\ref{fig:constraints}.

At low masses, $m_\phi \lesssim 1/T$, this picture breaks down. In this limit, the signal no longer oscillates over the observing time and looks like a polynomial expanded in $m_\phi t$. PTAs are not sensitive to the first few terms, up to $\mathcal{O}(t^2)$, in this polynomial since they are degenerate with the timing model~\cite{Ramani:2020hdo}. The first term they are sensitive to is $\propto A (m_\phi t)^3$, and therefore at low masses, $\text{SNR} \propto A (m_\phi T)^3$ and the constraints weaken as $d \propto 1/m_\phi$. In addition to these subtraction effects, red noise can also deteriorate the signal significance~\cite{Lee:2021zqw} at low masses. This explains why the constraints begin to flatten before turning to the $d \propto 1/m_\phi$ scaling. 

Lastly, we note that the reduced sensitivity around $f\sim \text{year}^{-1}$ is due to the fact that the relative position of the pulsar, which oscillates yearly, is not known to high precision~\cite{Edwards:2006zg,Hobbs:2006cd} and must be fit away, therefore reducing sensitivity.

\subsection{Connections to Other ULDM Effects in PTAs}
\label{subsec:connection_to_similar_effects}
\setlength{\tabcolsep}{12pt}
\begin{center}
    \begin{table*}[ht]
        \begin{tabular}{@{}cccccc@{}}
            \toprule\addlinespace
            Effect & $A$ & P$/$E & $\omega = 2 \pi f$ & $\langle h_I h_J \rangle$ & Refs. \\\addlinespace
            \cmidrule(lr){1-6} \\
            Shapiro & $\displaystyle\frac{\pi G \rho_\phi}{2 m_\phi^3}$ & P$+$E & $\displaystyle 2m_\phi$ & 1 & \cite{Khmelnitsky:2013lxt, Kato:2019bqz, Porayko:2014rfa, Porayko:2018sfa} \\ \addlinespace
            Doppler & \textcolor{BrickRed}{$\displaystyle ( \vec{y} \cdot \vec{d} ) \frac{\sqrt{2\rho_\phi}}{m_\phi^2 \Lambda} v_\phi$} & P$+$E & $\displaystyle m_\phi$ & $\cos{\theta_{IJ}}$ & - \\\addlinespace
            Pulsar Spin Fluctuations & $\displaystyle ( \vec{y} \cdot \vec{d} )\frac{\sqrt{2\rho_\phi}}{m_\phi^2 \Lambda}$ & P & $\displaystyle m_\phi$ & $\delta_{IJ}$ & Eq.~\eqref{eq:pulsar_spin_fluctuation_signal}\\ \addlinespace
            Reference Clock Shift & $\displaystyle ( \vec{y} \cdot \vec{d} ) \frac{\sqrt{2\rho_\phi}}{m_\phi^2 \Lambda}$ & E & $\displaystyle m_\phi$ & $1$ & Eq.~\eqref{eq:reference_clock_shift_signal}\\ \addlinespace\bottomrule
        \end{tabular}
    \caption{Summary of effects in PTA timing residuals from scalar ULDM candidates. $A$ is the signal amplitude, P$/$E denotes whether there are ``Pulsar" or ``Earth" terms in the signal, assuming the signal in each pulsar is written as $h = A_P \sin(\omega + \gamma_P) + A_E \sin(\omega + \gamma_E)$. Note that the sensitivity parameters, $\vec{y}$, will depend on whether the term comes from a pulsar or Earth effect. $\omega$ is the signal frequency, and $\langle h_I h_J \rangle$ represents the signal correlation between pulsars. Effects which are velocity suppressed are highlighted in \textcolor{BrickRed}{red}.}
    \label{tab:effect_summary}
    \end{table*}
\end{center}

We will conclude this section by discussing the specific differences in the ULDM induced signals presented here versus those considered previously~\cite{Khmelnitsky:2013lxt, Kato:2019bqz, Porayko:2014rfa, Porayko:2018sfa, Pierce:2018xmy, PPTA:2021uzb}. A summary of our discussion is given in Table~\ref{tab:effect_summary}, previous effects are labeled ``Shapiro" and ``Doppler", and the new effects are labeled ``Pulsar Spin Fluctuations" and ``Reference Clock Shift". The primary purpose of most PTAs is to measure the stochastic gravitational wave background. These ripples in space-time affect the photons' geodesic when traveling from the pulsar to the Earth. ULDM can induce a similar effect; the local ULDM density sources fluctuations to the metric which can affect light travel time in near complete analogy with the gravitational waves PTAs are searching for. This effect only depends on the background ULDM density and is model independent; any ULDM model which constitutes all of the DM will produce this effect. The amplitude of the signal is given by,
\begin{align}
    A_\mathrm{grav} = \frac{\pi G \rho_\phi}{2 m_\phi^2} \, ,
    \label{eq:A_grav}
\end{align}
see Ref.~\cite{Khmelnitsky:2013lxt} for more details of its derivation. To justify ignoring it in our previous analysis we require that the amplitude of the effects considered here is greater than the amplitude in Eq.~\eqref{eq:A_grav}.
This happens when
\begin{align}
    d_i \gtrsim \frac{4.5 \times 10^{-9}}{y_i} \left( \frac{10^{-23}\,\text{eV}}{m_\phi} \right) \, ,
\end{align}
and explains the gray shaded region in the bottom left corner of Figs.~\ref{fig:constraints}. In this region the purely gravitational signal is stronger and a more thorough analysis should include all effects. However these signals are not degenerate for a few reasons. First, the purely gravitational signal has a frequency of $\omega = 2 m_\phi$ whereas the model dependent ULDM signals have a frequency of $\omega = m_\phi$. In addition, the pulsar spin fluctuation signal is uncorrelated between pulsars, whereas the purely gravitational one is perfectly correlated giving another method of disentangling the signals.

Recently, the Doppler effect from vector ULDM has been studied using the Parkes PTA dataset~\cite{PPTA:2021uzb}. This effect can be understood by thinking about the vector DM sourcing a dark ``electric" field from $\partial_t \boldsymbol{\phi}$. If the pulsar or Earth are charged under this dark force they will accelerate in the presence of this field. A similar effect is present for scalar ULDM, however the acceleration induced will be due to the spatial derivative of the field, $a \sim \nabla \phi$, leading to a suppression since $v_\phi \sim 10^{-3}$. Therefore this effect will be subdominant to those discussed here which are not velocity suppressed. However the signal correlations are dipolar between pulsars which could potentially be used to search for this smaller signal. 

\section{Conclusions}
\label{sec:conclusions}

While the primary goal of PTAs is to detect the stochastic gravitational wave background (SGWB), their remarkable sensitivity can be leveraged for new physics searches. From searching for novel contributions to the GWB produced via cosmic strings~\cite{Siemens:2006yp,Blanco-Pillado:2017rnf}, inflationary fluctuations~\cite{Grishchuk1975,Lasky:2015lej,Vagnozzi:2020gtf}, and cosmological phase transitions~\cite{NANOGrav:2021flc,Winicour1973,Hogan1986,Deryagin:1986qq,Caprini:2010xv,Kobakhidze:2017mru} to signals produced from passing DM substructure~\cite{Ramani:2020hdo,Lee:2021zqw, Lee:2020wfn, Dror:2019twh} PTAs have been shown to set quite powerful constraints. In this work we have discussed two types of signals generically expected from scalar ULDM models with direct couplings to the SM. These couplings will generate apparent fluctuations in the SM fundamental constants, e.g., particle masses and couplings, which are then realized as new contributions to the timing residuals.

The two signals we discussed were: ``pulsar spin fluctuations", Sec.~\ref{subsec:pulsar_spin_fluctuations}, generated via changes to the moment of inertia and conservation of angular momentum, and ``reference clock shifts", Sec.~\ref{subsec:reference_clock_shifts} which change the tick frequency of the reference clocks used for the PTA. Notably, neither of these effects suffer from a velocity suppression as a Doppler shift would. We find that for a wide range of ULDM models the IPTA, and future PTA, can compete or outperform other constraints in the $10^{-23} \, \text{eV} \lesssim m_\phi \lesssim 10^{-20} \, \text{eV}$ mass window such as other atomic clock experiments, torsion balances, and the orbital decay of neutron star binaries. 

Despite the fact that the results derived in this paper were obtained using mock data, the formalism and code employed are ready to be deployed on real PTA datasets. Having proved the constraining power of PTAs for these kinds of ULDM models, we plan to apply the analyses performed in this paper to NANOGrav and IPTA data in the near future. 

\begin{acknowledgments}
    We thank Stephen Taylor for patiently helping us with \texttt{enterprise}, as well as giving valuable feedback on the manuscript. We also thank Jim Cordes for helpful comments, and Yufeng Du, Ryan Janish, and Vincent S. H. Lee for useful discussions.  D.K. was supported by the National Science Foundation under Grant No. PHY-1818899.  A.M. and T.T. were supported by the U.S. Department of Energy, Office of Science, Office of High Energy Physics, under Award No. DE-SC0021431, and the Quantum Information Science Enabled Discovery (QuantISED) for High Energy Physics (KA2401032). The computations presented here were conducted in the Resnick High Performance Computing Center, a facility supported by Resnick Sustainability Institute at the California Institute of Technology.
\end{acknowledgments}

\bibliographystyle{apsrev4-1}
\bibliography{bibliography}

\begin{thebibliography}{86}%
\makeatletter
\providecommand \@ifxundefined [1]{%
 \@ifx{#1\undefined}
}%
\providecommand \@ifnum [1]{%
 \ifnum #1\expandafter \@firstoftwo
 \else \expandafter \@secondoftwo
 \fi
}%
\providecommand \@ifx [1]{%
 \ifx #1\expandafter \@firstoftwo
 \else \expandafter \@secondoftwo
 \fi
}%
\providecommand \natexlab [1]{#1}%
\providecommand \enquote  [1]{``#1''}%
\providecommand \bibnamefont  [1]{#1}%
\providecommand \bibfnamefont [1]{#1}%
\providecommand \citenamefont [1]{#1}%
\providecommand \href@noop [0]{\@secondoftwo}%
\providecommand \href [0]{\begingroup \@sanitize@url \@href}%
\providecommand \@href[1]{\@@startlink{#1}\@@href}%
\providecommand \@@href[1]{\endgroup#1\@@endlink}%
\providecommand \@sanitize@url [0]{\catcode `\\12\catcode `\$12\catcode
  `\&12\catcode `\#12\catcode `\^12\catcode `\_12\catcode `\%12\relax}%
\providecommand \@@startlink[1]{}%
\providecommand \@@endlink[0]{}%
\providecommand \url  [0]{\begingroup\@sanitize@url \@url }%
\providecommand \@url [1]{\endgroup\@href {#1}{\urlprefix }}%
\providecommand \urlprefix  [0]{URL }%
\providecommand \Eprint [0]{\href }%
\providecommand \doibase [0]{http://dx.doi.org/}%
\providecommand \selectlanguage [0]{\@gobble}%
\providecommand \bibinfo  [0]{\@secondoftwo}%
\providecommand \bibfield  [0]{\@secondoftwo}%
\providecommand \translation [1]{[#1]}%
\providecommand \BibitemOpen [0]{}%
\providecommand \bibitemStop [0]{}%
\providecommand \bibitemNoStop [0]{.\EOS\space}%
\providecommand \EOS [0]{\spacefactor3000\relax}%
\providecommand \BibitemShut  [1]{\csname bibitem#1\endcsname}%
\let\auto@bib@innerbib\@empty
\bibitem [{\citenamefont {Tremaine}\ and\ \citenamefont
  {Gunn}(1979)}]{Tremaine:1979we}%
  \BibitemOpen
  \bibfield  {author} {\bibinfo {author} {\bibfnamefont {S.}~\bibnamefont
  {Tremaine}}\ and\ \bibinfo {author} {\bibfnamefont {J.~E.}\ \bibnamefont
  {Gunn}},\ }\href {\doibase 10.1103/PhysRevLett.42.407} {\bibfield  {journal}
  {\bibinfo  {journal} {Phys. Rev. Lett.}\ }\textbf {\bibinfo {volume} {42}},\
  \bibinfo {pages} {407} (\bibinfo {year} {1979})}\BibitemShut {NoStop}%
\bibitem [{\citenamefont {Di~Paolo}\ \emph {et~al.}(2018)\citenamefont
  {Di~Paolo}, \citenamefont {Nesti},\ and\ \citenamefont
  {Villante}}]{DiPaolo:2017geq}%
  \BibitemOpen
  \bibfield  {author} {\bibinfo {author} {\bibfnamefont {C.}~\bibnamefont
  {Di~Paolo}}, \bibinfo {author} {\bibfnamefont {F.}~\bibnamefont {Nesti}}, \
  and\ \bibinfo {author} {\bibfnamefont {F.~L.}\ \bibnamefont {Villante}},\
  }\href {\doibase 10.1093/mnras/sty091} {\bibfield  {journal} {\bibinfo
  {journal} {Mon. Not. Roy. Astron. Soc.}\ }\textbf {\bibinfo {volume} {475}},\
  \bibinfo {pages} {5385} (\bibinfo {year} {2018})},\ \Eprint
  {http://arxiv.org/abs/1704.06644} {arXiv:1704.06644 [astro-ph.GA]}
  \BibitemShut {NoStop}%
\bibitem [{\citenamefont {Savchenko}\ and\ \citenamefont
  {Rudakovskyi}(2019)}]{Savchenko:2019qnn}%
  \BibitemOpen
  \bibfield  {author} {\bibinfo {author} {\bibfnamefont {D.}~\bibnamefont
  {Savchenko}}\ and\ \bibinfo {author} {\bibfnamefont {A.}~\bibnamefont
  {Rudakovskyi}},\ }\href {\doibase 10.1093/mnras/stz1573} {\bibfield
  {journal} {\bibinfo  {journal} {Mon. Not. Roy. Astron. Soc.}\ }\textbf
  {\bibinfo {volume} {487}},\ \bibinfo {pages} {5711} (\bibinfo {year}
  {2019})},\ \Eprint {http://arxiv.org/abs/1903.01862} {arXiv:1903.01862
  [astro-ph.CO]} \BibitemShut {NoStop}%
\bibitem [{\citenamefont {Alvey}\ \emph {et~al.}(2021)\citenamefont {Alvey},
  \citenamefont {Sabti}, \citenamefont {Tiki}, \citenamefont {Blas},
  \citenamefont {Bondarenko}, \citenamefont {Boyarsky}, \citenamefont
  {Escudero}, \citenamefont {Fairbairn}, \citenamefont {Orkney},\ and\
  \citenamefont {Read}}]{Alvey:2020xsk}%
  \BibitemOpen
  \bibfield  {author} {\bibinfo {author} {\bibfnamefont {J.}~\bibnamefont
  {Alvey}}, \bibinfo {author} {\bibfnamefont {N.}~\bibnamefont {Sabti}},
  \bibinfo {author} {\bibfnamefont {V.}~\bibnamefont {Tiki}}, \bibinfo {author}
  {\bibfnamefont {D.}~\bibnamefont {Blas}}, \bibinfo {author} {\bibfnamefont
  {K.}~\bibnamefont {Bondarenko}}, \bibinfo {author} {\bibfnamefont
  {A.}~\bibnamefont {Boyarsky}}, \bibinfo {author} {\bibfnamefont
  {M.}~\bibnamefont {Escudero}}, \bibinfo {author} {\bibfnamefont
  {M.}~\bibnamefont {Fairbairn}}, \bibinfo {author} {\bibfnamefont
  {M.}~\bibnamefont {Orkney}}, \ and\ \bibinfo {author} {\bibfnamefont {J.~I.}\
  \bibnamefont {Read}},\ }\href {\doibase 10.1093/mnras/staa3640} {\bibfield
  {journal} {\bibinfo  {journal} {Mon. Not. Roy. Astron. Soc.}\ }\textbf
  {\bibinfo {volume} {501}},\ \bibinfo {pages} {1188} (\bibinfo {year}
  {2021})},\ \Eprint {http://arxiv.org/abs/2010.03572} {arXiv:2010.03572
  [hep-ph]} \BibitemShut {NoStop}%
\bibitem [{\citenamefont {Armengaud}\ \emph {et~al.}(2017)\citenamefont
  {Armengaud}, \citenamefont {Palanque-Delabrouille}, \citenamefont {Y\`eche},
  \citenamefont {Marsh},\ and\ \citenamefont {Baur}}]{Armengaud:2017nkf}%
  \BibitemOpen
  \bibfield  {author} {\bibinfo {author} {\bibfnamefont {E.}~\bibnamefont
  {Armengaud}}, \bibinfo {author} {\bibfnamefont {N.}~\bibnamefont
  {Palanque-Delabrouille}}, \bibinfo {author} {\bibfnamefont {C.}~\bibnamefont
  {Y\`eche}}, \bibinfo {author} {\bibfnamefont {D.~J.~E.}\ \bibnamefont
  {Marsh}}, \ and\ \bibinfo {author} {\bibfnamefont {J.}~\bibnamefont {Baur}},\
  }\href {\doibase 10.1093/mnras/stx1870} {\bibfield  {journal} {\bibinfo
  {journal} {Mon. Not. Roy. Astron. Soc.}\ }\textbf {\bibinfo {volume} {471}},\
  \bibinfo {pages} {4606} (\bibinfo {year} {2017})},\ \Eprint
  {http://arxiv.org/abs/1703.09126} {arXiv:1703.09126 [astro-ph.CO]}
  \BibitemShut {NoStop}%
\bibitem [{\citenamefont {Bullock}\ and\ \citenamefont
  {Boylan-Kolchin}(2017)}]{Bullock:2017xww}%
  \BibitemOpen
  \bibfield  {author} {\bibinfo {author} {\bibfnamefont {J.~S.}\ \bibnamefont
  {Bullock}}\ and\ \bibinfo {author} {\bibfnamefont {M.}~\bibnamefont
  {Boylan-Kolchin}},\ }\href {\doibase 10.1146/annurev-astro-091916-055313}
  {\bibfield  {journal} {\bibinfo  {journal} {Ann. Rev. Astron. Astrophys.}\
  }\textbf {\bibinfo {volume} {55}},\ \bibinfo {pages} {343} (\bibinfo {year}
  {2017})},\ \Eprint {http://arxiv.org/abs/1707.04256} {arXiv:1707.04256
  [astro-ph.CO]} \BibitemShut {NoStop}%
\bibitem [{\citenamefont {Arvanitaki}\ \emph {et~al.}(2015)\citenamefont
  {Arvanitaki}, \citenamefont {Huang},\ and\ \citenamefont
  {Van~Tilburg}}]{Arvanitaki:2014faa}%
  \BibitemOpen
  \bibfield  {author} {\bibinfo {author} {\bibfnamefont {A.}~\bibnamefont
  {Arvanitaki}}, \bibinfo {author} {\bibfnamefont {J.}~\bibnamefont {Huang}}, \
  and\ \bibinfo {author} {\bibfnamefont {K.}~\bibnamefont {Van~Tilburg}},\
  }\href {\doibase 10.1103/PhysRevD.91.015015} {\bibfield  {journal} {\bibinfo
  {journal} {Phys. Rev. D}\ }\textbf {\bibinfo {volume} {91}},\ \bibinfo
  {pages} {015015} (\bibinfo {year} {2015})},\ \Eprint
  {http://arxiv.org/abs/1405.2925} {arXiv:1405.2925 [hep-ph]} \BibitemShut
  {NoStop}%
\bibitem [{\citenamefont {Damour}\ and\ \citenamefont
  {Donoghue}(2010{\natexlab{a}})}]{Damour:2010rp}%
  \BibitemOpen
  \bibfield  {author} {\bibinfo {author} {\bibfnamefont {T.}~\bibnamefont
  {Damour}}\ and\ \bibinfo {author} {\bibfnamefont {J.~F.}\ \bibnamefont
  {Donoghue}},\ }\href {\doibase 10.1103/PhysRevD.82.084033} {\bibfield
  {journal} {\bibinfo  {journal} {Phys. Rev. D}\ }\textbf {\bibinfo {volume}
  {82}},\ \bibinfo {pages} {084033} (\bibinfo {year} {2010}{\natexlab{a}})},\
  \Eprint {http://arxiv.org/abs/1007.2792} {arXiv:1007.2792 [gr-qc]}
  \BibitemShut {NoStop}%
\bibitem [{\citenamefont {Hui}\ \emph {et~al.}(2017)\citenamefont {Hui},
  \citenamefont {Ostriker}, \citenamefont {Tremaine},\ and\ \citenamefont
  {Witten}}]{Hui:2016ltb}%
  \BibitemOpen
  \bibfield  {author} {\bibinfo {author} {\bibfnamefont {L.}~\bibnamefont
  {Hui}}, \bibinfo {author} {\bibfnamefont {J.~P.}\ \bibnamefont {Ostriker}},
  \bibinfo {author} {\bibfnamefont {S.}~\bibnamefont {Tremaine}}, \ and\
  \bibinfo {author} {\bibfnamefont {E.}~\bibnamefont {Witten}},\ }\href
  {\doibase 10.1103/PhysRevD.95.043541} {\bibfield  {journal} {\bibinfo
  {journal} {Phys. Rev. D}\ }\textbf {\bibinfo {volume} {95}},\ \bibinfo
  {pages} {043541} (\bibinfo {year} {2017})},\ \Eprint
  {http://arxiv.org/abs/1610.08297} {arXiv:1610.08297 [astro-ph.CO]}
  \BibitemShut {NoStop}%
\bibitem [{\citenamefont {Damour}\ and\ \citenamefont
  {Donoghue}(2010{\natexlab{b}})}]{Damour:2010rm}%
  \BibitemOpen
  \bibfield  {author} {\bibinfo {author} {\bibfnamefont {T.}~\bibnamefont
  {Damour}}\ and\ \bibinfo {author} {\bibfnamefont {J.~F.}\ \bibnamefont
  {Donoghue}},\ }\href {\doibase 10.1088/0264-9381/27/20/202001} {\bibfield
  {journal} {\bibinfo  {journal} {Class. Quant. Grav.}\ }\textbf {\bibinfo
  {volume} {27}},\ \bibinfo {pages} {202001} (\bibinfo {year}
  {2010}{\natexlab{b}})},\ \Eprint {http://arxiv.org/abs/1007.2790}
  {arXiv:1007.2790 [gr-qc]} \BibitemShut {NoStop}%
\bibitem [{\citenamefont {Hu}\ \emph {et~al.}(2000)\citenamefont {Hu},
  \citenamefont {Barkana},\ and\ \citenamefont {Gruzinov}}]{Hu:2000ke}%
  \BibitemOpen
  \bibfield  {author} {\bibinfo {author} {\bibfnamefont {W.}~\bibnamefont
  {Hu}}, \bibinfo {author} {\bibfnamefont {R.}~\bibnamefont {Barkana}}, \ and\
  \bibinfo {author} {\bibfnamefont {A.}~\bibnamefont {Gruzinov}},\ }\href
  {\doibase 10.1103/PhysRevLett.85.1158} {\bibfield  {journal} {\bibinfo
  {journal} {Phys. Rev. Lett.}\ }\textbf {\bibinfo {volume} {85}},\ \bibinfo
  {pages} {1158} (\bibinfo {year} {2000})},\ \Eprint
  {http://arxiv.org/abs/astro-ph/0003365} {arXiv:astro-ph/0003365} \BibitemShut
  {NoStop}%
\bibitem [{\citenamefont {Marsh}(2016)}]{Marsh:2015xka}%
  \BibitemOpen
  \bibfield  {author} {\bibinfo {author} {\bibfnamefont {D.~J.~E.}\
  \bibnamefont {Marsh}},\ }\href {\doibase 10.1016/j.physrep.2016.06.005}
  {\bibfield  {journal} {\bibinfo  {journal} {Phys. Rept.}\ }\textbf {\bibinfo
  {volume} {643}},\ \bibinfo {pages} {1} (\bibinfo {year} {2016})},\ \Eprint
  {http://arxiv.org/abs/1510.07633} {arXiv:1510.07633 [astro-ph.CO]}
  \BibitemShut {NoStop}%
\bibitem [{\citenamefont {Arkani-Hamed}\ \emph {et~al.}(2003)\citenamefont
  {Arkani-Hamed}, \citenamefont {Georgi},\ and\ \citenamefont
  {Schwartz}}]{Arkani-Hamed:2002bjr}%
  \BibitemOpen
  \bibfield  {author} {\bibinfo {author} {\bibfnamefont {N.}~\bibnamefont
  {Arkani-Hamed}}, \bibinfo {author} {\bibfnamefont {H.}~\bibnamefont
  {Georgi}}, \ and\ \bibinfo {author} {\bibfnamefont {M.~D.}\ \bibnamefont
  {Schwartz}},\ }\href {\doibase 10.1016/S0003-4916(03)00068-X} {\bibfield
  {journal} {\bibinfo  {journal} {Annals Phys.}\ }\textbf {\bibinfo {volume}
  {305}},\ \bibinfo {pages} {96} (\bibinfo {year} {2003})},\ \Eprint
  {http://arxiv.org/abs/hep-th/0210184} {arXiv:hep-th/0210184} \BibitemShut
  {NoStop}%
\bibitem [{\citenamefont {Centers}\ \emph {et~al.}(2021)\citenamefont {Centers}
  \emph {et~al.}}]{Centers:2019dyn}%
  \BibitemOpen
  \bibfield  {author} {\bibinfo {author} {\bibfnamefont {G.~P.}\ \bibnamefont
  {Centers}} \emph {et~al.},\ }\href {\doibase 10.1038/s41467-021-27632-7}
  {\bibfield  {journal} {\bibinfo  {journal} {Nature Commun.}\ }\textbf
  {\bibinfo {volume} {12}},\ \bibinfo {pages} {7321} (\bibinfo {year}
  {2021})},\ \Eprint {http://arxiv.org/abs/1905.13650} {arXiv:1905.13650
  [astro-ph.CO]} \BibitemShut {NoStop}%
\bibitem [{\citenamefont {Hlozek}\ \emph {et~al.}(2015)\citenamefont {Hlozek},
  \citenamefont {Grin}, \citenamefont {Marsh},\ and\ \citenamefont
  {Ferreira}}]{Hlozek:2014lca}%
  \BibitemOpen
  \bibfield  {author} {\bibinfo {author} {\bibfnamefont {R.}~\bibnamefont
  {Hlozek}}, \bibinfo {author} {\bibfnamefont {D.}~\bibnamefont {Grin}},
  \bibinfo {author} {\bibfnamefont {D.~J.~E.}\ \bibnamefont {Marsh}}, \ and\
  \bibinfo {author} {\bibfnamefont {P.~G.}\ \bibnamefont {Ferreira}},\ }\href
  {\doibase 10.1103/PhysRevD.91.103512} {\bibfield  {journal} {\bibinfo
  {journal} {Phys. Rev. D}\ }\textbf {\bibinfo {volume} {91}},\ \bibinfo
  {pages} {103512} (\bibinfo {year} {2015})},\ \Eprint
  {http://arxiv.org/abs/1410.2896} {arXiv:1410.2896 [astro-ph.CO]} \BibitemShut
  {NoStop}%
\bibitem [{\citenamefont {Hlo\v{z}ek}\ \emph {et~al.}(2017)\citenamefont
  {Hlo\v{z}ek}, \citenamefont {Marsh}, \citenamefont {Grin}, \citenamefont
  {Allison}, \citenamefont {Dunkley},\ and\ \citenamefont
  {Calabrese}}]{Hlozek:2016lzm}%
  \BibitemOpen
  \bibfield  {author} {\bibinfo {author} {\bibfnamefont {R.}~\bibnamefont
  {Hlo\v{z}ek}}, \bibinfo {author} {\bibfnamefont {D.~J.~E.}\ \bibnamefont
  {Marsh}}, \bibinfo {author} {\bibfnamefont {D.}~\bibnamefont {Grin}},
  \bibinfo {author} {\bibfnamefont {R.}~\bibnamefont {Allison}}, \bibinfo
  {author} {\bibfnamefont {J.}~\bibnamefont {Dunkley}}, \ and\ \bibinfo
  {author} {\bibfnamefont {E.}~\bibnamefont {Calabrese}},\ }\href {\doibase
  10.1103/PhysRevD.95.123511} {\bibfield  {journal} {\bibinfo  {journal} {Phys.
  Rev. D}\ }\textbf {\bibinfo {volume} {95}},\ \bibinfo {pages} {123511}
  (\bibinfo {year} {2017})},\ \Eprint {http://arxiv.org/abs/1607.08208}
  {arXiv:1607.08208 [astro-ph.CO]} \BibitemShut {NoStop}%
\bibitem [{\citenamefont {Hlozek}\ \emph {et~al.}(2018)\citenamefont {Hlozek},
  \citenamefont {Marsh},\ and\ \citenamefont {Grin}}]{Hlozek:2017zzf}%
  \BibitemOpen
  \bibfield  {author} {\bibinfo {author} {\bibfnamefont {R.}~\bibnamefont
  {Hlozek}}, \bibinfo {author} {\bibfnamefont {D.~J.~E.}\ \bibnamefont
  {Marsh}}, \ and\ \bibinfo {author} {\bibfnamefont {D.}~\bibnamefont {Grin}},\
  }\href {\doibase 10.1093/mnras/sty271} {\bibfield  {journal} {\bibinfo
  {journal} {Mon. Not. Roy. Astron. Soc.}\ }\textbf {\bibinfo {volume} {476}},\
  \bibinfo {pages} {3063} (\bibinfo {year} {2018})},\ \Eprint
  {http://arxiv.org/abs/1708.05681} {arXiv:1708.05681 [astro-ph.CO]}
  \BibitemShut {NoStop}%
\bibitem [{\citenamefont {Schive}\ \emph
  {et~al.}(2014{\natexlab{a}})\citenamefont {Schive}, \citenamefont {Chiueh},\
  and\ \citenamefont {Broadhurst}}]{Schive:2014dra}%
  \BibitemOpen
  \bibfield  {author} {\bibinfo {author} {\bibfnamefont {H.-Y.}\ \bibnamefont
  {Schive}}, \bibinfo {author} {\bibfnamefont {T.}~\bibnamefont {Chiueh}}, \
  and\ \bibinfo {author} {\bibfnamefont {T.}~\bibnamefont {Broadhurst}},\
  }\href {\doibase 10.1038/nphys2996} {\bibfield  {journal} {\bibinfo
  {journal} {Nature Phys.}\ }\textbf {\bibinfo {volume} {10}},\ \bibinfo
  {pages} {496} (\bibinfo {year} {2014}{\natexlab{a}})},\ \Eprint
  {http://arxiv.org/abs/1406.6586} {arXiv:1406.6586 [astro-ph.GA]} \BibitemShut
  {NoStop}%
\bibitem [{\citenamefont {Schive}\ \emph
  {et~al.}(2014{\natexlab{b}})\citenamefont {Schive}, \citenamefont {Liao},
  \citenamefont {Woo}, \citenamefont {Wong}, \citenamefont {Chiueh},
  \citenamefont {Broadhurst},\ and\ \citenamefont {Hwang}}]{Schive:2014hza}%
  \BibitemOpen
  \bibfield  {author} {\bibinfo {author} {\bibfnamefont {H.-Y.}\ \bibnamefont
  {Schive}}, \bibinfo {author} {\bibfnamefont {M.-H.}\ \bibnamefont {Liao}},
  \bibinfo {author} {\bibfnamefont {T.-P.}\ \bibnamefont {Woo}}, \bibinfo
  {author} {\bibfnamefont {S.-K.}\ \bibnamefont {Wong}}, \bibinfo {author}
  {\bibfnamefont {T.}~\bibnamefont {Chiueh}}, \bibinfo {author} {\bibfnamefont
  {T.}~\bibnamefont {Broadhurst}}, \ and\ \bibinfo {author} {\bibfnamefont
  {W.~Y.~P.}\ \bibnamefont {Hwang}},\ }\href {\doibase
  10.1103/PhysRevLett.113.261302} {\bibfield  {journal} {\bibinfo  {journal}
  {Phys. Rev. Lett.}\ }\textbf {\bibinfo {volume} {113}},\ \bibinfo {pages}
  {261302} (\bibinfo {year} {2014}{\natexlab{b}})},\ \Eprint
  {http://arxiv.org/abs/1407.7762} {arXiv:1407.7762 [astro-ph.GA]} \BibitemShut
  {NoStop}%
\bibitem [{\citenamefont {Bar}\ \emph {et~al.}(2018)\citenamefont {Bar},
  \citenamefont {Blas}, \citenamefont {Blum},\ and\ \citenamefont
  {Sibiryakov}}]{Bar:2018acw}%
  \BibitemOpen
  \bibfield  {author} {\bibinfo {author} {\bibfnamefont {N.}~\bibnamefont
  {Bar}}, \bibinfo {author} {\bibfnamefont {D.}~\bibnamefont {Blas}}, \bibinfo
  {author} {\bibfnamefont {K.}~\bibnamefont {Blum}}, \ and\ \bibinfo {author}
  {\bibfnamefont {S.}~\bibnamefont {Sibiryakov}},\ }\href {\doibase
  10.1103/PhysRevD.98.083027} {\bibfield  {journal} {\bibinfo  {journal} {Phys.
  Rev. D}\ }\textbf {\bibinfo {volume} {98}},\ \bibinfo {pages} {083027}
  (\bibinfo {year} {2018})},\ \Eprint {http://arxiv.org/abs/1805.00122}
  {arXiv:1805.00122 [astro-ph.CO]} \BibitemShut {NoStop}%
\bibitem [{\citenamefont {Bar}\ \emph {et~al.}(2022)\citenamefont {Bar},
  \citenamefont {Blum},\ and\ \citenamefont {Sun}}]{Bar:2021kti}%
  \BibitemOpen
  \bibfield  {author} {\bibinfo {author} {\bibfnamefont {N.}~\bibnamefont
  {Bar}}, \bibinfo {author} {\bibfnamefont {K.}~\bibnamefont {Blum}}, \ and\
  \bibinfo {author} {\bibfnamefont {C.}~\bibnamefont {Sun}},\ }\href {\doibase
  10.1103/PhysRevD.105.083015} {\bibfield  {journal} {\bibinfo  {journal}
  {Phys. Rev. D}\ }\textbf {\bibinfo {volume} {105}},\ \bibinfo {pages}
  {083015} (\bibinfo {year} {2022})},\ \Eprint
  {http://arxiv.org/abs/2111.03070} {arXiv:2111.03070 [hep-ph]} \BibitemShut
  {NoStop}%
\bibitem [{\citenamefont {Nadler}\ \emph {et~al.}(2019)\citenamefont {Nadler},
  \citenamefont {Gluscevic}, \citenamefont {Boddy},\ and\ \citenamefont
  {Wechsler}}]{Nadler:2019zrb}%
  \BibitemOpen
  \bibfield  {author} {\bibinfo {author} {\bibfnamefont {E.~O.}\ \bibnamefont
  {Nadler}}, \bibinfo {author} {\bibfnamefont {V.}~\bibnamefont {Gluscevic}},
  \bibinfo {author} {\bibfnamefont {K.~K.}\ \bibnamefont {Boddy}}, \ and\
  \bibinfo {author} {\bibfnamefont {R.~H.}\ \bibnamefont {Wechsler}},\ }\href
  {\doibase 10.3847/2041-8213/ab1eb2} {\bibfield  {journal} {\bibinfo
  {journal} {Astrophys. J. Lett.}\ }\textbf {\bibinfo {volume} {878}},\
  \bibinfo {pages} {32} (\bibinfo {year} {2019})},\ \bibinfo {note} {[Erratum:
  Astrophys.J.Lett. 897, L46 (2020), Erratum: Astrophys.J. 897, L46 (2020)]},\
  \Eprint {http://arxiv.org/abs/1904.10000} {arXiv:1904.10000 [astro-ph.CO]}
  \BibitemShut {NoStop}%
\bibitem [{\citenamefont {Schutz}(2020)}]{Schutz:2020jox}%
  \BibitemOpen
  \bibfield  {author} {\bibinfo {author} {\bibfnamefont {K.}~\bibnamefont
  {Schutz}},\ }\href {\doibase 10.1103/PhysRevD.101.123026} {\bibfield
  {journal} {\bibinfo  {journal} {Phys. Rev. D}\ }\textbf {\bibinfo {volume}
  {101}},\ \bibinfo {pages} {123026} (\bibinfo {year} {2020})},\ \Eprint
  {http://arxiv.org/abs/2001.05503} {arXiv:2001.05503 [astro-ph.CO]}
  \BibitemShut {NoStop}%
\bibitem [{\citenamefont {Ir\v{s}i\v{c}}\ \emph {et~al.}(2017)\citenamefont
  {Ir\v{s}i\v{c}}, \citenamefont {Viel}, \citenamefont {Haehnelt},
  \citenamefont {Bolton},\ and\ \citenamefont {Becker}}]{Irsic:2017yje}%
  \BibitemOpen
  \bibfield  {author} {\bibinfo {author} {\bibfnamefont {V.}~\bibnamefont
  {Ir\v{s}i\v{c}}}, \bibinfo {author} {\bibfnamefont {M.}~\bibnamefont {Viel}},
  \bibinfo {author} {\bibfnamefont {M.~G.}\ \bibnamefont {Haehnelt}}, \bibinfo
  {author} {\bibfnamefont {J.~S.}\ \bibnamefont {Bolton}}, \ and\ \bibinfo
  {author} {\bibfnamefont {G.~D.}\ \bibnamefont {Becker}},\ }\href {\doibase
  10.1103/PhysRevLett.119.031302} {\bibfield  {journal} {\bibinfo  {journal}
  {Phys. Rev. Lett.}\ }\textbf {\bibinfo {volume} {119}},\ \bibinfo {pages}
  {031302} (\bibinfo {year} {2017})},\ \Eprint
  {http://arxiv.org/abs/1703.04683} {arXiv:1703.04683 [astro-ph.CO]}
  \BibitemShut {NoStop}%
\bibitem [{\citenamefont {Nori}\ \emph {et~al.}(2019)\citenamefont {Nori},
  \citenamefont {Murgia}, \citenamefont {Ir\v{s}i\v{c}}, \citenamefont
  {Baldi},\ and\ \citenamefont {Viel}}]{Nori:2018pka}%
  \BibitemOpen
  \bibfield  {author} {\bibinfo {author} {\bibfnamefont {M.}~\bibnamefont
  {Nori}}, \bibinfo {author} {\bibfnamefont {R.}~\bibnamefont {Murgia}},
  \bibinfo {author} {\bibfnamefont {V.}~\bibnamefont {Ir\v{s}i\v{c}}}, \bibinfo
  {author} {\bibfnamefont {M.}~\bibnamefont {Baldi}}, \ and\ \bibinfo {author}
  {\bibfnamefont {M.}~\bibnamefont {Viel}},\ }\href {\doibase
  10.1093/mnras/sty2888} {\bibfield  {journal} {\bibinfo  {journal} {Mon. Not.
  Roy. Astron. Soc.}\ }\textbf {\bibinfo {volume} {482}},\ \bibinfo {pages}
  {3227} (\bibinfo {year} {2019})},\ \Eprint {http://arxiv.org/abs/1809.09619}
  {arXiv:1809.09619 [astro-ph.CO]} \BibitemShut {NoStop}%
\bibitem [{\citenamefont {Khmelnitsky}\ and\ \citenamefont
  {Rubakov}(2014)}]{Khmelnitsky:2013lxt}%
  \BibitemOpen
  \bibfield  {author} {\bibinfo {author} {\bibfnamefont {A.}~\bibnamefont
  {Khmelnitsky}}\ and\ \bibinfo {author} {\bibfnamefont {V.}~\bibnamefont
  {Rubakov}},\ }\href {\doibase 10.1088/1475-7516/2014/02/019} {\bibfield
  {journal} {\bibinfo  {journal} {JCAP}\ }\textbf {\bibinfo {volume} {02}},\
  \bibinfo {pages} {019} (\bibinfo {year} {2014})},\ \Eprint
  {http://arxiv.org/abs/1309.5888} {arXiv:1309.5888 [astro-ph.CO]} \BibitemShut
  {NoStop}%
\bibitem [{\citenamefont {Kato}\ and\ \citenamefont
  {Soda}(2020)}]{Kato:2019bqz}%
  \BibitemOpen
  \bibfield  {author} {\bibinfo {author} {\bibfnamefont {R.}~\bibnamefont
  {Kato}}\ and\ \bibinfo {author} {\bibfnamefont {J.}~\bibnamefont {Soda}},\
  }\href {\doibase 10.1088/1475-7516/2020/09/036} {\bibfield  {journal}
  {\bibinfo  {journal} {JCAP}\ }\textbf {\bibinfo {volume} {09}},\ \bibinfo
  {pages} {036} (\bibinfo {year} {2020})},\ \Eprint
  {http://arxiv.org/abs/1904.09143} {arXiv:1904.09143 [astro-ph.HE]}
  \BibitemShut {NoStop}%
\bibitem [{\citenamefont {Porayko}\ and\ \citenamefont
  {Postnov}(2014)}]{Porayko:2014rfa}%
  \BibitemOpen
  \bibfield  {author} {\bibinfo {author} {\bibfnamefont {N.~K.}\ \bibnamefont
  {Porayko}}\ and\ \bibinfo {author} {\bibfnamefont {K.~A.}\ \bibnamefont
  {Postnov}},\ }\href {\doibase 10.1103/PhysRevD.90.062008} {\bibfield
  {journal} {\bibinfo  {journal} {Phys. Rev. D}\ }\textbf {\bibinfo {volume}
  {90}},\ \bibinfo {pages} {062008} (\bibinfo {year} {2014})},\ \Eprint
  {http://arxiv.org/abs/1408.4670} {arXiv:1408.4670 [astro-ph.CO]} \BibitemShut
  {NoStop}%
\bibitem [{\citenamefont {Porayko}\ \emph {et~al.}(2018)\citenamefont {Porayko}
  \emph {et~al.}}]{Porayko:2018sfa}%
  \BibitemOpen
  \bibfield  {author} {\bibinfo {author} {\bibfnamefont {N.~K.}\ \bibnamefont
  {Porayko}} \emph {et~al.},\ }\href {\doibase 10.1103/PhysRevD.98.102002}
  {\bibfield  {journal} {\bibinfo  {journal} {Phys. Rev. D}\ }\textbf {\bibinfo
  {volume} {98}},\ \bibinfo {pages} {102002} (\bibinfo {year} {2018})},\
  \Eprint {http://arxiv.org/abs/1810.03227} {arXiv:1810.03227 [astro-ph.CO]}
  \BibitemShut {NoStop}%
\bibitem [{\citenamefont {Kennedy}\ \emph {et~al.}(2020)\citenamefont
  {Kennedy}, \citenamefont {Oelker}, \citenamefont {Robinson}, \citenamefont
  {Bothwell}, \citenamefont {Kedar}, \citenamefont {Milner}, \citenamefont
  {Marti}, \citenamefont {Derevianko},\ and\ \citenamefont
  {Ye}}]{Kennedy:2020bac}%
  \BibitemOpen
  \bibfield  {author} {\bibinfo {author} {\bibfnamefont {C.~J.}\ \bibnamefont
  {Kennedy}}, \bibinfo {author} {\bibfnamefont {E.}~\bibnamefont {Oelker}},
  \bibinfo {author} {\bibfnamefont {J.~M.}\ \bibnamefont {Robinson}}, \bibinfo
  {author} {\bibfnamefont {T.}~\bibnamefont {Bothwell}}, \bibinfo {author}
  {\bibfnamefont {D.}~\bibnamefont {Kedar}}, \bibinfo {author} {\bibfnamefont
  {W.~R.}\ \bibnamefont {Milner}}, \bibinfo {author} {\bibfnamefont {G.~E.}\
  \bibnamefont {Marti}}, \bibinfo {author} {\bibfnamefont {A.}~\bibnamefont
  {Derevianko}}, \ and\ \bibinfo {author} {\bibfnamefont {J.}~\bibnamefont
  {Ye}},\ }\href {\doibase 10.1103/PhysRevLett.125.201302} {\bibfield
  {journal} {\bibinfo  {journal} {Phys. Rev. Lett.}\ }\textbf {\bibinfo
  {volume} {125}},\ \bibinfo {pages} {201302} (\bibinfo {year} {2020})},\
  \Eprint {http://arxiv.org/abs/2008.08773} {arXiv:2008.08773
  [physics.atom-ph]} \BibitemShut {NoStop}%
\bibitem [{\citenamefont {Hees}\ \emph {et~al.}(2016)\citenamefont {Hees},
  \citenamefont {Gu\'ena}, \citenamefont {Abgrall}, \citenamefont {Bize},\ and\
  \citenamefont {Wolf}}]{Hees:2016gop}%
  \BibitemOpen
  \bibfield  {author} {\bibinfo {author} {\bibfnamefont {A.}~\bibnamefont
  {Hees}}, \bibinfo {author} {\bibfnamefont {J.}~\bibnamefont {Gu\'ena}},
  \bibinfo {author} {\bibfnamefont {M.}~\bibnamefont {Abgrall}}, \bibinfo
  {author} {\bibfnamefont {S.}~\bibnamefont {Bize}}, \ and\ \bibinfo {author}
  {\bibfnamefont {P.}~\bibnamefont {Wolf}},\ }\href {\doibase
  10.1103/PhysRevLett.117.061301} {\bibfield  {journal} {\bibinfo  {journal}
  {Phys. Rev. Lett.}\ }\textbf {\bibinfo {volume} {117}},\ \bibinfo {pages}
  {061301} (\bibinfo {year} {2016})},\ \Eprint
  {http://arxiv.org/abs/1604.08514} {arXiv:1604.08514 [gr-qc]} \BibitemShut
  {NoStop}%
\bibitem [{\citenamefont {Wcis\l{}o}\ \emph {et~al.}(2018)\citenamefont
  {Wcis\l{}o} \emph {et~al.}}]{Wcislo:2018ojh}%
  \BibitemOpen
  \bibfield  {author} {\bibinfo {author} {\bibfnamefont {P.}~\bibnamefont
  {Wcis\l{}o}} \emph {et~al.},\ }\href {\doibase 10.1126/sciadv.aau4869}
  {\bibfield  {journal} {\bibinfo  {journal} {Sci. Adv.}\ }\textbf {\bibinfo
  {volume} {4}},\ \bibinfo {pages} {eaau4869} (\bibinfo {year} {2018})},\
  \Eprint {http://arxiv.org/abs/1806.04762} {arXiv:1806.04762
  [physics.atom-ph]} \BibitemShut {NoStop}%
\bibitem [{\citenamefont {Van~Tilburg}\ \emph {et~al.}(2015)\citenamefont
  {Van~Tilburg}, \citenamefont {Leefer}, \citenamefont {Bougas},\ and\
  \citenamefont {Budker}}]{VanTilburg:2015oza}%
  \BibitemOpen
  \bibfield  {author} {\bibinfo {author} {\bibfnamefont {K.}~\bibnamefont
  {Van~Tilburg}}, \bibinfo {author} {\bibfnamefont {N.}~\bibnamefont {Leefer}},
  \bibinfo {author} {\bibfnamefont {L.}~\bibnamefont {Bougas}}, \ and\ \bibinfo
  {author} {\bibfnamefont {D.}~\bibnamefont {Budker}},\ }\href {\doibase
  10.1103/PhysRevLett.115.011802} {\bibfield  {journal} {\bibinfo  {journal}
  {Phys. Rev. Lett.}\ }\textbf {\bibinfo {volume} {115}},\ \bibinfo {pages}
  {011802} (\bibinfo {year} {2015})},\ \Eprint
  {http://arxiv.org/abs/1503.06886} {arXiv:1503.06886 [physics.atom-ph]}
  \BibitemShut {NoStop}%
\bibitem [{\citenamefont {Flambaum}\ \emph {et~al.}(2004)\citenamefont
  {Flambaum}, \citenamefont {Leinweber}, \citenamefont {Thomas},\ and\
  \citenamefont {Young}}]{Flambaum:2004tm}%
  \BibitemOpen
  \bibfield  {author} {\bibinfo {author} {\bibfnamefont {V.~V.}\ \bibnamefont
  {Flambaum}}, \bibinfo {author} {\bibfnamefont {D.~B.}\ \bibnamefont
  {Leinweber}}, \bibinfo {author} {\bibfnamefont {A.~W.}\ \bibnamefont
  {Thomas}}, \ and\ \bibinfo {author} {\bibfnamefont {R.~D.}\ \bibnamefont
  {Young}},\ }\href {\doibase 10.1103/PhysRevD.69.115006} {\bibfield  {journal}
  {\bibinfo  {journal} {Phys. Rev. D}\ }\textbf {\bibinfo {volume} {69}},\
  \bibinfo {pages} {115006} (\bibinfo {year} {2004})},\ \Eprint
  {http://arxiv.org/abs/hep-ph/0402098} {arXiv:hep-ph/0402098} \BibitemShut
  {NoStop}%
\bibitem [{\citenamefont {Network}\ \emph {et~al.}(2020)\citenamefont
  {Network}, \citenamefont {Collaboration}, \citenamefont {Beloy},
  \citenamefont {Bodine}, \citenamefont {Bothwell}, \citenamefont {Brewer},
  \citenamefont {Bromley}, \citenamefont {Chen}, \citenamefont {Deschênes},
  \citenamefont {Diddams}, \citenamefont {Fasano}, \citenamefont {Fortier},
  \citenamefont {Hassan}, \citenamefont {Hume}, \citenamefont {Kedar},
  \citenamefont {Kennedy}, \citenamefont {Khader}, \citenamefont {Koepke},
  \citenamefont {Leibrandt}, \citenamefont {Leopardi}, \citenamefont {Ludlow},
  \citenamefont {McGrew}, \citenamefont {Milner}, \citenamefont {Newbury},
  \citenamefont {Nicolodi}, \citenamefont {Oelker}, \citenamefont {Parker},
  \citenamefont {Robinson}, \citenamefont {Romisch}, \citenamefont {Schäffer},
  \citenamefont {Sherman}, \citenamefont {Sinclair}, \citenamefont
  {Sonderhouse}, \citenamefont {Swann}, \citenamefont {Yao}, \citenamefont
  {Ye},\ and\ \citenamefont {Zhang}}]{Network2020}%
  \BibitemOpen
  \bibfield  {author} {\bibinfo {author} {\bibfnamefont {B.~A. C.~O.}\
  \bibnamefont {Network}}, \bibinfo {author} {\bibnamefont {Collaboration}},
  \bibinfo {author} {\bibfnamefont {K.}~\bibnamefont {Beloy}}, \bibinfo
  {author} {\bibfnamefont {M.~I.}\ \bibnamefont {Bodine}}, \bibinfo {author}
  {\bibfnamefont {T.}~\bibnamefont {Bothwell}}, \bibinfo {author}
  {\bibfnamefont {S.~M.}\ \bibnamefont {Brewer}}, \bibinfo {author}
  {\bibfnamefont {S.~L.}\ \bibnamefont {Bromley}}, \bibinfo {author}
  {\bibfnamefont {J.-S.}\ \bibnamefont {Chen}}, \bibinfo {author}
  {\bibfnamefont {J.-D.}\ \bibnamefont {Deschênes}}, \bibinfo {author}
  {\bibfnamefont {S.~A.}\ \bibnamefont {Diddams}}, \bibinfo {author}
  {\bibfnamefont {R.~J.}\ \bibnamefont {Fasano}}, \bibinfo {author}
  {\bibfnamefont {T.~M.}\ \bibnamefont {Fortier}}, \bibinfo {author}
  {\bibfnamefont {Y.~S.}\ \bibnamefont {Hassan}}, \bibinfo {author}
  {\bibfnamefont {D.~B.}\ \bibnamefont {Hume}}, \bibinfo {author}
  {\bibfnamefont {D.}~\bibnamefont {Kedar}}, \bibinfo {author} {\bibfnamefont
  {C.~J.}\ \bibnamefont {Kennedy}}, \bibinfo {author} {\bibfnamefont
  {I.}~\bibnamefont {Khader}}, \bibinfo {author} {\bibfnamefont
  {A.}~\bibnamefont {Koepke}}, \bibinfo {author} {\bibfnamefont {D.~R.}\
  \bibnamefont {Leibrandt}}, \bibinfo {author} {\bibfnamefont {H.}~\bibnamefont
  {Leopardi}}, \bibinfo {author} {\bibfnamefont {A.~D.}\ \bibnamefont
  {Ludlow}}, \bibinfo {author} {\bibfnamefont {W.~F.}\ \bibnamefont {McGrew}},
  \bibinfo {author} {\bibfnamefont {W.~R.}\ \bibnamefont {Milner}}, \bibinfo
  {author} {\bibfnamefont {N.~R.}\ \bibnamefont {Newbury}}, \bibinfo {author}
  {\bibfnamefont {D.}~\bibnamefont {Nicolodi}}, \bibinfo {author}
  {\bibfnamefont {E.}~\bibnamefont {Oelker}}, \bibinfo {author} {\bibfnamefont
  {T.~E.}\ \bibnamefont {Parker}}, \bibinfo {author} {\bibfnamefont {J.~M.}\
  \bibnamefont {Robinson}}, \bibinfo {author} {\bibfnamefont {S.}~\bibnamefont
  {Romisch}}, \bibinfo {author} {\bibfnamefont {S.~A.}\ \bibnamefont
  {Schäffer}}, \bibinfo {author} {\bibfnamefont {J.~A.}\ \bibnamefont
  {Sherman}}, \bibinfo {author} {\bibfnamefont {L.~C.}\ \bibnamefont
  {Sinclair}}, \bibinfo {author} {\bibfnamefont {L.}~\bibnamefont
  {Sonderhouse}}, \bibinfo {author} {\bibfnamefont {W.~C.}\ \bibnamefont
  {Swann}}, \bibinfo {author} {\bibfnamefont {J.}~\bibnamefont {Yao}}, \bibinfo
  {author} {\bibfnamefont {J.}~\bibnamefont {Ye}}, \ and\ \bibinfo {author}
  {\bibfnamefont {X.}~\bibnamefont {Zhang}},\ }\href
  {http://arxiv.org/abs/2005.14694} {\bibfield  {journal} {\bibinfo  {journal}
  {arXiv:2005.14694 [physics]}\ } (\bibinfo {year} {2020})},\ \bibinfo {note}
  {arXiv: 2005.14694}\BibitemShut {NoStop}%
\bibitem [{\citenamefont {Graham}\ \emph {et~al.}(2016)\citenamefont {Graham},
  \citenamefont {Kaplan}, \citenamefont {Mardon}, \citenamefont {Rajendran},\
  and\ \citenamefont {Terrano}}]{Graham:2015ifn}%
  \BibitemOpen
  \bibfield  {author} {\bibinfo {author} {\bibfnamefont {P.~W.}\ \bibnamefont
  {Graham}}, \bibinfo {author} {\bibfnamefont {D.~E.}\ \bibnamefont {Kaplan}},
  \bibinfo {author} {\bibfnamefont {J.}~\bibnamefont {Mardon}}, \bibinfo
  {author} {\bibfnamefont {S.}~\bibnamefont {Rajendran}}, \ and\ \bibinfo
  {author} {\bibfnamefont {W.~A.}\ \bibnamefont {Terrano}},\ }\href {\doibase
  10.1103/PhysRevD.93.075029} {\bibfield  {journal} {\bibinfo  {journal} {Phys.
  Rev. D}\ }\textbf {\bibinfo {volume} {93}},\ \bibinfo {pages} {075029}
  (\bibinfo {year} {2016})},\ \Eprint {http://arxiv.org/abs/1512.06165}
  {arXiv:1512.06165 [hep-ph]} \BibitemShut {NoStop}%
\bibitem [{\citenamefont {Morisaki}\ \emph {et~al.}(2021)\citenamefont
  {Morisaki}, \citenamefont {Fujita}, \citenamefont {Michimura}, \citenamefont
  {Nakatsuka},\ and\ \citenamefont {Obata}}]{Morisaki:2020gui}%
  \BibitemOpen
  \bibfield  {author} {\bibinfo {author} {\bibfnamefont {S.}~\bibnamefont
  {Morisaki}}, \bibinfo {author} {\bibfnamefont {T.}~\bibnamefont {Fujita}},
  \bibinfo {author} {\bibfnamefont {Y.}~\bibnamefont {Michimura}}, \bibinfo
  {author} {\bibfnamefont {H.}~\bibnamefont {Nakatsuka}}, \ and\ \bibinfo
  {author} {\bibfnamefont {I.}~\bibnamefont {Obata}},\ }\href {\doibase
  10.1103/PhysRevD.103.L051702} {\bibfield  {journal} {\bibinfo  {journal}
  {Phys. Rev. D}\ }\textbf {\bibinfo {volume} {103}},\ \bibinfo {pages}
  {L051702} (\bibinfo {year} {2021})},\ \Eprint
  {http://arxiv.org/abs/2011.03589} {arXiv:2011.03589 [hep-ph]} \BibitemShut
  {NoStop}%
\bibitem [{\citenamefont {Pierce}\ \emph {et~al.}(2018)\citenamefont {Pierce},
  \citenamefont {Riles},\ and\ \citenamefont {Zhao}}]{Pierce:2018xmy}%
  \BibitemOpen
  \bibfield  {author} {\bibinfo {author} {\bibfnamefont {A.}~\bibnamefont
  {Pierce}}, \bibinfo {author} {\bibfnamefont {K.}~\bibnamefont {Riles}}, \
  and\ \bibinfo {author} {\bibfnamefont {Y.}~\bibnamefont {Zhao}},\ }\href
  {\doibase 10.1103/PhysRevLett.121.061102} {\bibfield  {journal} {\bibinfo
  {journal} {Phys. Rev. Lett.}\ }\textbf {\bibinfo {volume} {121}},\ \bibinfo
  {pages} {061102} (\bibinfo {year} {2018})},\ \Eprint
  {http://arxiv.org/abs/1801.10161} {arXiv:1801.10161 [hep-ph]} \BibitemShut
  {NoStop}%
\bibitem [{\citenamefont {Xue}\ \emph {et~al.}(2022)\citenamefont {Xue} \emph
  {et~al.}}]{PPTA:2021uzb}%
  \BibitemOpen
  \bibfield  {author} {\bibinfo {author} {\bibfnamefont {X.}~\bibnamefont
  {Xue}} \emph {et~al.} (\bibinfo {collaboration} {PPTA}),\ }\href {\doibase
  10.1103/PhysRevResearch.4.L012022} {\bibfield  {journal} {\bibinfo  {journal}
  {Phys. Rev. Res.}\ }\textbf {\bibinfo {volume} {4}},\ \bibinfo {pages}
  {L012022} (\bibinfo {year} {2022})},\ \bibinfo {note} {arXiv: 2112.07687},\
  \Eprint {http://arxiv.org/abs/2112.07687} {arXiv:2112.07687 [hep-ph]}
  \BibitemShut {NoStop}%
\bibitem [{\citenamefont {Blas}\ \emph {et~al.}(2017)\citenamefont {Blas},
  \citenamefont {Nacir},\ and\ \citenamefont {Sibiryakov}}]{Blas:2016ddr}%
  \BibitemOpen
  \bibfield  {author} {\bibinfo {author} {\bibfnamefont {D.}~\bibnamefont
  {Blas}}, \bibinfo {author} {\bibfnamefont {D.~L.}\ \bibnamefont {Nacir}}, \
  and\ \bibinfo {author} {\bibfnamefont {S.}~\bibnamefont {Sibiryakov}},\
  }\href {\doibase 10.1103/PhysRevLett.118.261102} {\bibfield  {journal}
  {\bibinfo  {journal} {Phys. Rev. Lett.}\ }\textbf {\bibinfo {volume} {118}},\
  \bibinfo {pages} {261102} (\bibinfo {year} {2017})},\ \Eprint
  {http://arxiv.org/abs/1612.06789} {arXiv:1612.06789 [hep-ph]} \BibitemShut
  {NoStop}%
\bibitem [{\citenamefont {Dror}\ \emph {et~al.}(2020)\citenamefont {Dror},
  \citenamefont {Laha},\ and\ \citenamefont {Opferkuch}}]{Dror:2019uea}%
  \BibitemOpen
  \bibfield  {author} {\bibinfo {author} {\bibfnamefont {J.~A.}\ \bibnamefont
  {Dror}}, \bibinfo {author} {\bibfnamefont {R.}~\bibnamefont {Laha}}, \ and\
  \bibinfo {author} {\bibfnamefont {T.}~\bibnamefont {Opferkuch}},\ }\href
  {\doibase 10.1103/PhysRevD.102.023005} {\bibfield  {journal} {\bibinfo
  {journal} {Phys. Rev. D}\ }\textbf {\bibinfo {volume} {102}},\ \bibinfo
  {pages} {023005} (\bibinfo {year} {2020})},\ \Eprint
  {http://arxiv.org/abs/1909.12845} {arXiv:1909.12845 [hep-ph]} \BibitemShut
  {NoStop}%
\bibitem [{\citenamefont {Kumar~Poddar}\ \emph {et~al.}(2019)\citenamefont
  {Kumar~Poddar}, \citenamefont {Mohanty},\ and\ \citenamefont
  {Jana}}]{KumarPoddar:2019ceq}%
  \BibitemOpen
  \bibfield  {author} {\bibinfo {author} {\bibfnamefont {T.}~\bibnamefont
  {Kumar~Poddar}}, \bibinfo {author} {\bibfnamefont {S.}~\bibnamefont
  {Mohanty}}, \ and\ \bibinfo {author} {\bibfnamefont {S.}~\bibnamefont
  {Jana}},\ }\href {\doibase 10.1103/PhysRevD.100.123023} {\bibfield  {journal}
  {\bibinfo  {journal} {Phys. Rev. D}\ }\textbf {\bibinfo {volume} {100}},\
  \bibinfo {pages} {123023} (\bibinfo {year} {2019})},\ \Eprint
  {http://arxiv.org/abs/1908.09732} {arXiv:1908.09732 [hep-ph]} \BibitemShut
  {NoStop}%
\bibitem [{\citenamefont {Schlamminger}\ \emph {et~al.}(2008)\citenamefont
  {Schlamminger}, \citenamefont {Choi}, \citenamefont {Wagner}, \citenamefont
  {Gundlach},\ and\ \citenamefont {Adelberger}}]{Schlamminger:2007ht}%
  \BibitemOpen
  \bibfield  {author} {\bibinfo {author} {\bibfnamefont {S.}~\bibnamefont
  {Schlamminger}}, \bibinfo {author} {\bibfnamefont {K.~Y.}\ \bibnamefont
  {Choi}}, \bibinfo {author} {\bibfnamefont {T.~A.}\ \bibnamefont {Wagner}},
  \bibinfo {author} {\bibfnamefont {J.~H.}\ \bibnamefont {Gundlach}}, \ and\
  \bibinfo {author} {\bibfnamefont {E.~G.}\ \bibnamefont {Adelberger}},\ }\href
  {\doibase 10.1103/PhysRevLett.100.041101} {\bibfield  {journal} {\bibinfo
  {journal} {Phys. Rev. Lett.}\ }\textbf {\bibinfo {volume} {100}},\ \bibinfo
  {pages} {041101} (\bibinfo {year} {2008})},\ \Eprint
  {http://arxiv.org/abs/0712.0607} {arXiv:0712.0607 [gr-qc]} \BibitemShut
  {NoStop}%
\bibitem [{\citenamefont {Berg\'e}\ \emph {et~al.}(2018)\citenamefont
  {Berg\'e}, \citenamefont {Brax}, \citenamefont {M\'etris}, \citenamefont
  {Pernot-Borr\`as}, \citenamefont {Touboul},\ and\ \citenamefont
  {Uzan}}]{Berge:2017ovy}%
  \BibitemOpen
  \bibfield  {author} {\bibinfo {author} {\bibfnamefont {J.}~\bibnamefont
  {Berg\'e}}, \bibinfo {author} {\bibfnamefont {P.}~\bibnamefont {Brax}},
  \bibinfo {author} {\bibfnamefont {G.}~\bibnamefont {M\'etris}}, \bibinfo
  {author} {\bibfnamefont {M.}~\bibnamefont {Pernot-Borr\`as}}, \bibinfo
  {author} {\bibfnamefont {P.}~\bibnamefont {Touboul}}, \ and\ \bibinfo
  {author} {\bibfnamefont {J.-P.}\ \bibnamefont {Uzan}},\ }\href {\doibase
  10.1103/PhysRevLett.120.141101} {\bibfield  {journal} {\bibinfo  {journal}
  {Phys. Rev. Lett.}\ }\textbf {\bibinfo {volume} {120}},\ \bibinfo {pages}
  {141101} (\bibinfo {year} {2018})},\ \Eprint
  {http://arxiv.org/abs/1712.00483} {arXiv:1712.00483 [gr-qc]} \BibitemShut
  {NoStop}%
\bibitem [{\citenamefont {Zhang}\ \emph {et~al.}(2019)\citenamefont {Zhang},
  \citenamefont {Liu},\ and\ \citenamefont {Chu}}]{Zhang2019}%
  \BibitemOpen
  \bibfield  {author} {\bibinfo {author} {\bibfnamefont {J.}~\bibnamefont
  {Zhang}}, \bibinfo {author} {\bibfnamefont {H.}~\bibnamefont {Liu}}, \ and\
  \bibinfo {author} {\bibfnamefont {M.-C.}\ \bibnamefont {Chu}},\ }\href
  {https://www.frontiersin.org/article/10.3389/fspas.2018.00048} {\bibfield
  {journal} {\bibinfo  {journal} {Frontiers in Astronomy and Space Sciences}\
  }\textbf {\bibinfo {volume} {5}} (\bibinfo {year} {2019})}\BibitemShut
  {NoStop}%
\bibitem [{\citenamefont {Press}\ and\ \citenamefont
  {Schechter}(1974)}]{Press:1973iz}%
  \BibitemOpen
  \bibfield  {author} {\bibinfo {author} {\bibfnamefont {W.~H.}\ \bibnamefont
  {Press}}\ and\ \bibinfo {author} {\bibfnamefont {P.}~\bibnamefont
  {Schechter}},\ }\href {\doibase 10.1086/152650} {\bibfield  {journal}
  {\bibinfo  {journal} {Astrophys. J.}\ }\textbf {\bibinfo {volume} {187}},\
  \bibinfo {pages} {425} (\bibinfo {year} {1974})}\BibitemShut {NoStop}%
\bibitem [{\citenamefont {Lim}\ \emph {et~al.}(2019)\citenamefont {Lim},
  \citenamefont {Holt},\ and\ \citenamefont {Stahulak}}]{Lim:2018xne}%
  \BibitemOpen
  \bibfield  {author} {\bibinfo {author} {\bibfnamefont {Y.}~\bibnamefont
  {Lim}}, \bibinfo {author} {\bibfnamefont {J.~W.}\ \bibnamefont {Holt}}, \
  and\ \bibinfo {author} {\bibfnamefont {R.~J.}\ \bibnamefont {Stahulak}},\
  }\href {\doibase 10.1103/PhysRevC.100.035802} {\bibfield  {journal} {\bibinfo
   {journal} {Phys. Rev. C}\ }\textbf {\bibinfo {volume} {100}},\ \bibinfo
  {pages} {035802} (\bibinfo {year} {2019})},\ \Eprint
  {http://arxiv.org/abs/1810.10992} {arXiv:1810.10992 [nucl-th]} \BibitemShut
  {NoStop}%
\bibitem [{\citenamefont {Bell}\ \emph {et~al.}(2019)\citenamefont {Bell},
  \citenamefont {Busoni},\ and\ \citenamefont {Robles}}]{Bell:2019pyc}%
  \BibitemOpen
  \bibfield  {author} {\bibinfo {author} {\bibfnamefont {N.~F.}\ \bibnamefont
  {Bell}}, \bibinfo {author} {\bibfnamefont {G.}~\bibnamefont {Busoni}}, \ and\
  \bibinfo {author} {\bibfnamefont {S.}~\bibnamefont {Robles}},\ }\href
  {\doibase 10.1088/1475-7516/2019/06/054} {\bibfield  {journal} {\bibinfo
  {journal} {JCAP}\ }\textbf {\bibinfo {volume} {06}},\ \bibinfo {pages} {054}
  (\bibinfo {year} {2019})},\ \Eprint {http://arxiv.org/abs/1904.09803}
  {arXiv:1904.09803 [hep-ph]} \BibitemShut {NoStop}%
\bibitem [{\citenamefont {Arzoumanian}\ \emph {et~al.}(2016)\citenamefont
  {Arzoumanian} \emph {et~al.}}]{NANOGrav:2015aud}%
  \BibitemOpen
  \bibfield  {author} {\bibinfo {author} {\bibfnamefont {Z.}~\bibnamefont
  {Arzoumanian}} \emph {et~al.} (\bibinfo {collaboration} {NANOGrav}),\ }\href
  {\doibase 10.3847/0004-637X/821/1/13} {\bibfield  {journal} {\bibinfo
  {journal} {Astrophys. J.}\ }\textbf {\bibinfo {volume} {821}},\ \bibinfo
  {pages} {13} (\bibinfo {year} {2016})},\ \Eprint
  {http://arxiv.org/abs/1508.03024} {arXiv:1508.03024 [astro-ph.GA]}
  \BibitemShut {NoStop}%
\bibitem [{\citenamefont {Taylor}(2021)}]{Taylor:2021yjx}%
  \BibitemOpen
  \bibfield  {author} {\bibinfo {author} {\bibfnamefont {S.~R.}\ \bibnamefont
  {Taylor}},\ }\href@noop {} {\  (\bibinfo {year} {2021})},\ \Eprint
  {http://arxiv.org/abs/2105.13270} {arXiv:2105.13270 [astro-ph.HE]}
  \BibitemShut {NoStop}%
\bibitem [{\citenamefont {Arzoumanian}\ \emph {et~al.}(2014)\citenamefont
  {Arzoumanian} \emph {et~al.}}]{NANOGrav:2014zwv}%
  \BibitemOpen
  \bibfield  {author} {\bibinfo {author} {\bibfnamefont {Z.}~\bibnamefont
  {Arzoumanian}} \emph {et~al.} (\bibinfo {collaboration} {NANOGrav}),\ }\href
  {\doibase 10.1088/0004-637X/794/2/141} {\bibfield  {journal} {\bibinfo
  {journal} {Astrophys. J.}\ }\textbf {\bibinfo {volume} {794}},\ \bibinfo
  {pages} {141} (\bibinfo {year} {2014})},\ \Eprint
  {http://arxiv.org/abs/1404.1267} {arXiv:1404.1267 [astro-ph.GA]} \BibitemShut
  {NoStop}%
\bibitem [{\citenamefont {Cordes}(2013)}]{Cordes:2013iea}%
  \BibitemOpen
  \bibfield  {author} {\bibinfo {author} {\bibfnamefont {J.~M.}\ \bibnamefont
  {Cordes}},\ }\href {\doibase 10.1088/0264-9381/30/22/224002} {\bibfield
  {journal} {\bibinfo  {journal} {Class. Quant. Grav.}\ }\textbf {\bibinfo
  {volume} {30}},\ \bibinfo {pages} {224002} (\bibinfo {year}
  {2013})}\BibitemShut {NoStop}%
\bibitem [{\citenamefont {Lentati}\ \emph {et~al.}(2013)\citenamefont
  {Lentati}, \citenamefont {Alexander}, \citenamefont {Hobson}, \citenamefont
  {Taylor}, \citenamefont {Gair}, \citenamefont {Balan},\ and\ \citenamefont
  {van Haasteren}}]{Lentati:2012xb}%
  \BibitemOpen
  \bibfield  {author} {\bibinfo {author} {\bibfnamefont {L.}~\bibnamefont
  {Lentati}}, \bibinfo {author} {\bibfnamefont {P.}~\bibnamefont {Alexander}},
  \bibinfo {author} {\bibfnamefont {M.~P.}\ \bibnamefont {Hobson}}, \bibinfo
  {author} {\bibfnamefont {S.}~\bibnamefont {Taylor}}, \bibinfo {author}
  {\bibfnamefont {J.}~\bibnamefont {Gair}}, \bibinfo {author} {\bibfnamefont
  {S.~T.}\ \bibnamefont {Balan}}, \ and\ \bibinfo {author} {\bibfnamefont
  {R.}~\bibnamefont {van Haasteren}},\ }\href {\doibase
  10.1103/PhysRevD.87.104021} {\bibfield  {journal} {\bibinfo  {journal} {Phys.
  Rev. D}\ }\textbf {\bibinfo {volume} {87}},\ \bibinfo {pages} {104021}
  (\bibinfo {year} {2013})},\ \Eprint {http://arxiv.org/abs/1210.3578}
  {arXiv:1210.3578 [astro-ph.IM]} \BibitemShut {NoStop}%
\bibitem [{\citenamefont {van Haasteren}\ and\ \citenamefont
  {Levin}(2013)}]{vanHaasteren:2012hj}%
  \BibitemOpen
  \bibfield  {author} {\bibinfo {author} {\bibfnamefont {R.}~\bibnamefont {van
  Haasteren}}\ and\ \bibinfo {author} {\bibfnamefont {Y.}~\bibnamefont
  {Levin}},\ }\href {\doibase 10.1093/mnras/sts097} {\bibfield  {journal}
  {\bibinfo  {journal} {Mon. Not. Roy. Astron. Soc.}\ }\textbf {\bibinfo
  {volume} {428}},\ \bibinfo {pages} {1147} (\bibinfo {year} {2013})},\ \Eprint
  {http://arxiv.org/abs/1202.5932} {arXiv:1202.5932 [astro-ph.IM]} \BibitemShut
  {NoStop}%
\bibitem [{\citenamefont {{Ellis}}\ and\ \citenamefont {{van
  Haasteren}}(2019)}]{2019ascl.soft12017E}%
  \BibitemOpen
  \bibfield  {author} {\bibinfo {author} {\bibfnamefont {J.}~\bibnamefont
  {{Ellis}}}\ and\ \bibinfo {author} {\bibfnamefont {R.}~\bibnamefont {{van
  Haasteren}}},\ }\href@noop {} {\enquote {\bibinfo {title} {{PTMCMCSampler:
  Parallel tempering MCMC sampler package written in Python}},}\ }\bibinfo
  {howpublished} {Astrophysics Source Code Library, record ascl:1912.017}
  (\bibinfo {year} {2019}),\ \Eprint {http://arxiv.org/abs/1912.017}
  {ascl:1912.017} \BibitemShut {NoStop}%
\bibitem [{\citenamefont {{Ellis}}\ \emph {et~al.}(2019)\citenamefont
  {{Ellis}}, \citenamefont {{Vallisneri}}, \citenamefont {{Taylor}},\ and\
  \citenamefont {{Baker}}}]{2019ascl.soft12015E}%
  \BibitemOpen
  \bibfield  {author} {\bibinfo {author} {\bibfnamefont {J.~A.}\ \bibnamefont
  {{Ellis}}}, \bibinfo {author} {\bibfnamefont {M.}~\bibnamefont
  {{Vallisneri}}}, \bibinfo {author} {\bibfnamefont {S.~R.}\ \bibnamefont
  {{Taylor}}}, \ and\ \bibinfo {author} {\bibfnamefont {P.~T.}\ \bibnamefont
  {{Baker}}},\ }\href@noop {} {\enquote {\bibinfo {title} {{ENTERPRISE:
  Enhanced Numerical Toolbox Enabling a Robust PulsaR Inference SuitE}},}\
  }\bibinfo {howpublished} {Astrophysics Source Code Library, record
  ascl:1912.015} (\bibinfo {year} {2019}),\ \Eprint
  {http://arxiv.org/abs/1912.015} {ascl:1912.015} \BibitemShut {NoStop}%
\bibitem [{\citenamefont {Taylor}\ \emph {et~al.}(2021)\citenamefont {Taylor},
  \citenamefont {Baker}, \citenamefont {Hazboun}, \citenamefont {Simon},\ and\
  \citenamefont {Vigeland}}]{enterprise}%
  \BibitemOpen
  \bibfield  {author} {\bibinfo {author} {\bibfnamefont {S.~R.}\ \bibnamefont
  {Taylor}}, \bibinfo {author} {\bibfnamefont {P.~T.}\ \bibnamefont {Baker}},
  \bibinfo {author} {\bibfnamefont {J.~S.}\ \bibnamefont {Hazboun}}, \bibinfo
  {author} {\bibfnamefont {J.}~\bibnamefont {Simon}}, \ and\ \bibinfo {author}
  {\bibfnamefont {S.~J.}\ \bibnamefont {Vigeland}},\ }\href
  {https://github.com/nanograv/enterprise_extensions} {\enquote {\bibinfo
  {title} {enterprise\_extensions},}\ } (\bibinfo {year} {2021}),\ \bibinfo
  {note} {v2.3.3}\BibitemShut {NoStop}%
\bibitem [{\citenamefont {Perera}\ \emph {et~al.}(2019)\citenamefont {Perera}
  \emph {et~al.}}]{Perera:2019sca}%
  \BibitemOpen
  \bibfield  {author} {\bibinfo {author} {\bibfnamefont {B.~B.~P.}\
  \bibnamefont {Perera}} \emph {et~al.},\ }\href {\doibase
  10.1093/mnras/stz2857} {\bibfield  {journal} {\bibinfo  {journal} {Mon. Not.
  Roy. Astron. Soc.}\ }\textbf {\bibinfo {volume} {490}},\ \bibinfo {pages}
  {4666} (\bibinfo {year} {2019})},\ \Eprint {http://arxiv.org/abs/1909.04534}
  {arXiv:1909.04534 [astro-ph.HE]} \BibitemShut {NoStop}%
\bibitem [{\citenamefont {Keane}\ \emph {et~al.}(2015)\citenamefont {Keane},
  \citenamefont {Bhattacharyya}, \citenamefont {Kramer}, \citenamefont
  {Stappers}, \citenamefont {Bates}, \citenamefont {Burgay}, \citenamefont
  {Chatterjee}, \citenamefont {Champion}, \citenamefont {Eatough},
  \citenamefont {Hessels}, \citenamefont {Janssen}, \citenamefont {Lee},
  \citenamefont {van Leeuwen}, \citenamefont {Margueron}, \citenamefont
  {Oertel}, \citenamefont {Possenti}, \citenamefont {Ransom}, \citenamefont
  {Theureau},\ and\ \citenamefont
  {Torne}}]{https://doi.org/10.48550/arxiv.1501.00056}%
  \BibitemOpen
  \bibfield  {author} {\bibinfo {author} {\bibfnamefont {E.~F.}\ \bibnamefont
  {Keane}}, \bibinfo {author} {\bibfnamefont {B.}~\bibnamefont
  {Bhattacharyya}}, \bibinfo {author} {\bibfnamefont {M.}~\bibnamefont
  {Kramer}}, \bibinfo {author} {\bibfnamefont {B.~W.}\ \bibnamefont
  {Stappers}}, \bibinfo {author} {\bibfnamefont {S.~D.}\ \bibnamefont {Bates}},
  \bibinfo {author} {\bibfnamefont {M.}~\bibnamefont {Burgay}}, \bibinfo
  {author} {\bibfnamefont {S.}~\bibnamefont {Chatterjee}}, \bibinfo {author}
  {\bibfnamefont {D.~J.}\ \bibnamefont {Champion}}, \bibinfo {author}
  {\bibfnamefont {R.~P.}\ \bibnamefont {Eatough}}, \bibinfo {author}
  {\bibfnamefont {J.~W.~T.}\ \bibnamefont {Hessels}}, \bibinfo {author}
  {\bibfnamefont {G.}~\bibnamefont {Janssen}}, \bibinfo {author} {\bibfnamefont
  {K.~J.}\ \bibnamefont {Lee}}, \bibinfo {author} {\bibfnamefont
  {J.}~\bibnamefont {van Leeuwen}}, \bibinfo {author} {\bibfnamefont
  {J.}~\bibnamefont {Margueron}}, \bibinfo {author} {\bibfnamefont
  {M.}~\bibnamefont {Oertel}}, \bibinfo {author} {\bibfnamefont
  {A.}~\bibnamefont {Possenti}}, \bibinfo {author} {\bibfnamefont
  {S.}~\bibnamefont {Ransom}}, \bibinfo {author} {\bibfnamefont
  {G.}~\bibnamefont {Theureau}}, \ and\ \bibinfo {author} {\bibfnamefont
  {P.}~\bibnamefont {Torne}},\ }\href {\doibase 10.48550/ARXIV.1501.00056}
  {\enquote {\bibinfo {title} {A cosmic census of radio pulsars with the
  ska},}\ } (\bibinfo {year} {2015})\BibitemShut {NoStop}%
\bibitem [{\citenamefont {Hallinan}\ \emph {et~al.}(2019)\citenamefont
  {Hallinan}, \citenamefont {Ravi}, \citenamefont {Weinreb}, \citenamefont
  {Kocz}, \citenamefont {Huang}, \citenamefont {Woody}, \citenamefont {Lamb},
  \citenamefont {D'Addario}, \citenamefont {Catha}, \citenamefont {Shi},
  \citenamefont {Law}, \citenamefont {Kulkarni}, \citenamefont {Phinney},
  \citenamefont {Eastwood}, \citenamefont {Bouman}, \citenamefont {McLaughlin},
  \citenamefont {Ransom}, \citenamefont {Siemens}, \citenamefont {Cordes},
  \citenamefont {Lynch}, \citenamefont {Kaplan}, \citenamefont {Chatterjee},
  \citenamefont {Lazio}, \citenamefont {Brazier}, \citenamefont {Bhatnagar},
  \citenamefont {Myers}, \citenamefont {Walter},\ and\ \citenamefont
  {Gaensler}}]{https://doi.org/10.48550/arxiv.1907.07648}%
  \BibitemOpen
  \bibfield  {author} {\bibinfo {author} {\bibfnamefont {G.}~\bibnamefont
  {Hallinan}}, \bibinfo {author} {\bibfnamefont {V.}~\bibnamefont {Ravi}},
  \bibinfo {author} {\bibfnamefont {S.}~\bibnamefont {Weinreb}}, \bibinfo
  {author} {\bibfnamefont {J.}~\bibnamefont {Kocz}}, \bibinfo {author}
  {\bibfnamefont {Y.}~\bibnamefont {Huang}}, \bibinfo {author} {\bibfnamefont
  {D.~P.}\ \bibnamefont {Woody}}, \bibinfo {author} {\bibfnamefont
  {J.}~\bibnamefont {Lamb}}, \bibinfo {author} {\bibfnamefont {L.}~\bibnamefont
  {D'Addario}}, \bibinfo {author} {\bibfnamefont {M.}~\bibnamefont {Catha}},
  \bibinfo {author} {\bibfnamefont {J.}~\bibnamefont {Shi}}, \bibinfo {author}
  {\bibfnamefont {C.}~\bibnamefont {Law}}, \bibinfo {author} {\bibfnamefont
  {S.~R.}\ \bibnamefont {Kulkarni}}, \bibinfo {author} {\bibfnamefont {E.~S.}\
  \bibnamefont {Phinney}}, \bibinfo {author} {\bibfnamefont {M.~W.}\
  \bibnamefont {Eastwood}}, \bibinfo {author} {\bibfnamefont {K.~L.}\
  \bibnamefont {Bouman}}, \bibinfo {author} {\bibfnamefont {M.~A.}\
  \bibnamefont {McLaughlin}}, \bibinfo {author} {\bibfnamefont {S.~M.}\
  \bibnamefont {Ransom}}, \bibinfo {author} {\bibfnamefont {X.}~\bibnamefont
  {Siemens}}, \bibinfo {author} {\bibfnamefont {J.~M.}\ \bibnamefont {Cordes}},
  \bibinfo {author} {\bibfnamefont {R.~S.}\ \bibnamefont {Lynch}}, \bibinfo
  {author} {\bibfnamefont {D.~L.}\ \bibnamefont {Kaplan}}, \bibinfo {author}
  {\bibfnamefont {S.}~\bibnamefont {Chatterjee}}, \bibinfo {author}
  {\bibfnamefont {J.}~\bibnamefont {Lazio}}, \bibinfo {author} {\bibfnamefont
  {A.}~\bibnamefont {Brazier}}, \bibinfo {author} {\bibfnamefont
  {S.}~\bibnamefont {Bhatnagar}}, \bibinfo {author} {\bibfnamefont {S.~T.}\
  \bibnamefont {Myers}}, \bibinfo {author} {\bibfnamefont {F.}~\bibnamefont
  {Walter}}, \ and\ \bibinfo {author} {\bibfnamefont {B.~M.}\ \bibnamefont
  {Gaensler}},\ }\href {\doibase 10.48550/ARXIV.1907.07648} {\enquote {\bibinfo
  {title} {The dsa-2000 -- a radio survey camera},}\ } (\bibinfo {year}
  {2019})\BibitemShut {NoStop}%
\bibitem [{\citenamefont {Hobbs}\ \emph {et~al.}(2019)\citenamefont {Hobbs},
  \citenamefont {Dai}, \citenamefont {Manchester}, \citenamefont {Shannon},
  \citenamefont {Kerr}, \citenamefont {Lee},\ and\ \citenamefont
  {Xu}}]{Hobbs:2014tqa}%
  \BibitemOpen
  \bibfield  {author} {\bibinfo {author} {\bibfnamefont {G.}~\bibnamefont
  {Hobbs}}, \bibinfo {author} {\bibfnamefont {S.}~\bibnamefont {Dai}}, \bibinfo
  {author} {\bibfnamefont {R.~N.}\ \bibnamefont {Manchester}}, \bibinfo
  {author} {\bibfnamefont {R.~M.}\ \bibnamefont {Shannon}}, \bibinfo {author}
  {\bibfnamefont {M.}~\bibnamefont {Kerr}}, \bibinfo {author} {\bibfnamefont
  {K.~J.}\ \bibnamefont {Lee}}, \ and\ \bibinfo {author} {\bibfnamefont
  {R.}~\bibnamefont {Xu}},\ }\href {\doibase 10.1088/1674-4527/19/2/20}
  {\bibfield  {journal} {\bibinfo  {journal} {Res. Astron. Astrophys.}\
  }\textbf {\bibinfo {volume} {19}},\ \bibinfo {pages} {020} (\bibinfo {year}
  {2019})},\ \Eprint {http://arxiv.org/abs/1407.0435} {arXiv:1407.0435
  [astro-ph.IM]} \BibitemShut {NoStop}%
\bibitem [{\citenamefont {Hazboun}\ \emph {et~al.}(2018)\citenamefont
  {Hazboun}, \citenamefont {Mingarelli},\ and\ \citenamefont
  {Lee}}]{Hazboun:2018wpv}%
  \BibitemOpen
  \bibfield  {author} {\bibinfo {author} {\bibfnamefont {J.~S.}\ \bibnamefont
  {Hazboun}}, \bibinfo {author} {\bibfnamefont {C.~M.~F.}\ \bibnamefont
  {Mingarelli}}, \ and\ \bibinfo {author} {\bibfnamefont {K.}~\bibnamefont
  {Lee}},\ }\href@noop {} {\  (\bibinfo {year} {2018})},\ \Eprint
  {http://arxiv.org/abs/1810.10527} {arXiv:1810.10527 [astro-ph.IM]}
  \BibitemShut {NoStop}%
\bibitem [{\citenamefont {{Vallisneri}}(2020)}]{2020ascl.soft02017V}%
  \BibitemOpen
  \bibfield  {author} {\bibinfo {author} {\bibfnamefont {M.}~\bibnamefont
  {{Vallisneri}}},\ }\href@noop {} {\enquote {\bibinfo {title} {{libstempo:
  Python wrapper for Tempo2}},}\ }\bibinfo {howpublished} {Astrophysics Source
  Code Library, record ascl:2002.017} (\bibinfo {year} {2020}),\ \Eprint
  {http://arxiv.org/abs/2002.017} {ascl:2002.017} \BibitemShut {NoStop}%
\bibitem [{\citenamefont {Hobbs}\ \emph {et~al.}(2006)\citenamefont {Hobbs},
  \citenamefont {Edwards},\ and\ \citenamefont {Manchester}}]{Hobbs:2006cd}%
  \BibitemOpen
  \bibfield  {author} {\bibinfo {author} {\bibfnamefont {G.}~\bibnamefont
  {Hobbs}}, \bibinfo {author} {\bibfnamefont {R.}~\bibnamefont {Edwards}}, \
  and\ \bibinfo {author} {\bibfnamefont {R.}~\bibnamefont {Manchester}},\
  }\href {\doibase 10.1111/j.1365-2966.2006.10302.x} {\bibfield  {journal}
  {\bibinfo  {journal} {Mon. Not. Roy. Astron. Soc.}\ }\textbf {\bibinfo
  {volume} {369}},\ \bibinfo {pages} {655} (\bibinfo {year} {2006})},\ \Eprint
  {http://arxiv.org/abs/astro-ph/0603381} {arXiv:astro-ph/0603381} \BibitemShut
  {NoStop}%
\bibitem [{\citenamefont {Edwards}\ \emph {et~al.}(2006)\citenamefont
  {Edwards}, \citenamefont {Hobbs},\ and\ \citenamefont
  {Manchester}}]{Edwards:2006zg}%
  \BibitemOpen
  \bibfield  {author} {\bibinfo {author} {\bibfnamefont {R.~T.}\ \bibnamefont
  {Edwards}}, \bibinfo {author} {\bibfnamefont {G.~B.}\ \bibnamefont {Hobbs}},
  \ and\ \bibinfo {author} {\bibfnamefont {R.~N.}\ \bibnamefont {Manchester}},\
  }\href {\doibase 10.1111/j.1365-2966.2006.10870.x} {\bibfield  {journal}
  {\bibinfo  {journal} {Mon. Not. Roy. Astron. Soc.}\ }\textbf {\bibinfo
  {volume} {372}},\ \bibinfo {pages} {1549} (\bibinfo {year} {2006})},\ \Eprint
  {http://arxiv.org/abs/astro-ph/0607664} {arXiv:astro-ph/0607664} \BibitemShut
  {NoStop}%
\bibitem [{\citenamefont {Janish}\ and\ \citenamefont
  {Ramani}(2020)}]{Janish:2020knz}%
  \BibitemOpen
  \bibfield  {author} {\bibinfo {author} {\bibfnamefont {R.}~\bibnamefont
  {Janish}}\ and\ \bibinfo {author} {\bibfnamefont {H.}~\bibnamefont
  {Ramani}},\ }\href {\doibase 10.1103/PhysRevD.102.115018} {\bibfield
  {journal} {\bibinfo  {journal} {Phys. Rev. D}\ }\textbf {\bibinfo {volume}
  {102}},\ \bibinfo {pages} {115018} (\bibinfo {year} {2020})},\ \Eprint
  {http://arxiv.org/abs/2006.10069} {arXiv:2006.10069 [hep-ph]} \BibitemShut
  {NoStop}%
\bibitem [{\citenamefont {Garani}\ and\ \citenamefont
  {Heeck}(2019)}]{Garani:2019fpa}%
  \BibitemOpen
  \bibfield  {author} {\bibinfo {author} {\bibfnamefont {R.}~\bibnamefont
  {Garani}}\ and\ \bibinfo {author} {\bibfnamefont {J.}~\bibnamefont {Heeck}},\
  }\href {\doibase 10.1103/PhysRevD.100.035039} {\bibfield  {journal} {\bibinfo
   {journal} {Phys. Rev. D}\ }\textbf {\bibinfo {volume} {100}},\ \bibinfo
  {pages} {035039} (\bibinfo {year} {2019})},\ \Eprint
  {http://arxiv.org/abs/1906.10145} {arXiv:1906.10145 [hep-ph]} \BibitemShut
  {NoStop}%
\bibitem [{\citenamefont {Zhang}\ and\ \citenamefont
  {Li}(2020)}]{Zhang:2020wov}%
  \BibitemOpen
  \bibfield  {author} {\bibinfo {author} {\bibfnamefont {N.-B.}\ \bibnamefont
  {Zhang}}\ and\ \bibinfo {author} {\bibfnamefont {B.-A.}\ \bibnamefont {Li}},\
  }\href {\doibase 10.3847/1538-4357/ab7dbc} {\bibfield  {journal} {\bibinfo
  {journal} {Astrophys. J.}\ }\textbf {\bibinfo {volume} {893}},\ \bibinfo
  {pages} {61} (\bibinfo {year} {2020})},\ \Eprint
  {http://arxiv.org/abs/2002.06446} {arXiv:2002.06446 [astro-ph.HE]}
  \BibitemShut {NoStop}%
\bibitem [{\citenamefont {Ramani}\ \emph {et~al.}(2020)\citenamefont {Ramani},
  \citenamefont {Trickle},\ and\ \citenamefont {Zurek}}]{Ramani:2020hdo}%
  \BibitemOpen
  \bibfield  {author} {\bibinfo {author} {\bibfnamefont {H.}~\bibnamefont
  {Ramani}}, \bibinfo {author} {\bibfnamefont {T.}~\bibnamefont {Trickle}}, \
  and\ \bibinfo {author} {\bibfnamefont {K.~M.}\ \bibnamefont {Zurek}},\ }\href
  {\doibase 10.1088/1475-7516/2020/12/033} {\bibfield  {journal} {\bibinfo
  {journal} {JCAP}\ }\textbf {\bibinfo {volume} {12}},\ \bibinfo {pages} {033}
  (\bibinfo {year} {2020})},\ \Eprint {http://arxiv.org/abs/2005.03030}
  {arXiv:2005.03030 [astro-ph.CO]} \BibitemShut {NoStop}%
\bibitem [{\citenamefont {Moore}\ \emph {et~al.}(2015)\citenamefont {Moore},
  \citenamefont {Cole},\ and\ \citenamefont {Berry}}]{Moore:2014lga}%
  \BibitemOpen
  \bibfield  {author} {\bibinfo {author} {\bibfnamefont {C.~J.}\ \bibnamefont
  {Moore}}, \bibinfo {author} {\bibfnamefont {R.~H.}\ \bibnamefont {Cole}}, \
  and\ \bibinfo {author} {\bibfnamefont {C.~P.~L.}\ \bibnamefont {Berry}},\
  }\href {\doibase 10.1088/0264-9381/32/1/015014} {\bibfield  {journal}
  {\bibinfo  {journal} {Class. Quant. Grav.}\ }\textbf {\bibinfo {volume}
  {32}},\ \bibinfo {pages} {015014} (\bibinfo {year} {2015})},\ \Eprint
  {http://arxiv.org/abs/1408.0740} {arXiv:1408.0740 [gr-qc]} \BibitemShut
  {NoStop}%
\bibitem [{\citenamefont {Smith}\ and\ \citenamefont
  {Caldwell}(2019)}]{Smith:2019wny}%
  \BibitemOpen
  \bibfield  {author} {\bibinfo {author} {\bibfnamefont {T.~L.}\ \bibnamefont
  {Smith}}\ and\ \bibinfo {author} {\bibfnamefont {R.}~\bibnamefont
  {Caldwell}},\ }\href {\doibase 10.1103/PhysRevD.100.104055} {\bibfield
  {journal} {\bibinfo  {journal} {Phys. Rev. D}\ }\textbf {\bibinfo {volume}
  {100}},\ \bibinfo {pages} {104055} (\bibinfo {year} {2019})},\ \Eprint
  {http://arxiv.org/abs/1908.00546} {arXiv:1908.00546 [astro-ph.CO]}
  \BibitemShut {NoStop}%
\bibitem [{\citenamefont {Allen}\ and\ \citenamefont
  {Romano}(1999)}]{Allen:1997ad}%
  \BibitemOpen
  \bibfield  {author} {\bibinfo {author} {\bibfnamefont {B.}~\bibnamefont
  {Allen}}\ and\ \bibinfo {author} {\bibfnamefont {J.~D.}\ \bibnamefont
  {Romano}},\ }\href {\doibase 10.1103/PhysRevD.59.102001} {\bibfield
  {journal} {\bibinfo  {journal} {Phys. Rev. D}\ }\textbf {\bibinfo {volume}
  {59}},\ \bibinfo {pages} {102001} (\bibinfo {year} {1999})},\ \Eprint
  {http://arxiv.org/abs/gr-qc/9710117} {arXiv:gr-qc/9710117} \BibitemShut
  {NoStop}%
\bibitem [{\citenamefont {Lee}\ \emph {et~al.}(2021{\natexlab{a}})\citenamefont
  {Lee}, \citenamefont {Taylor}, \citenamefont {Trickle},\ and\ \citenamefont
  {Zurek}}]{Lee:2021zqw}%
  \BibitemOpen
  \bibfield  {author} {\bibinfo {author} {\bibfnamefont {V.~S.~H.}\
  \bibnamefont {Lee}}, \bibinfo {author} {\bibfnamefont {S.~R.}\ \bibnamefont
  {Taylor}}, \bibinfo {author} {\bibfnamefont {T.}~\bibnamefont {Trickle}}, \
  and\ \bibinfo {author} {\bibfnamefont {K.~M.}\ \bibnamefont {Zurek}},\ }\href
  {\doibase 10.1088/1475-7516/2021/08/025} {\bibfield  {journal} {\bibinfo
  {journal} {JCAP}\ }\textbf {\bibinfo {volume} {08}},\ \bibinfo {pages} {025}
  (\bibinfo {year} {2021}{\natexlab{a}})},\ \Eprint
  {http://arxiv.org/abs/2104.05717} {arXiv:2104.05717 [astro-ph.CO]}
  \BibitemShut {NoStop}%
\bibitem [{\citenamefont {Siemens}\ \emph {et~al.}(2007)\citenamefont
  {Siemens}, \citenamefont {Mandic},\ and\ \citenamefont
  {Creighton}}]{Siemens:2006yp}%
  \BibitemOpen
  \bibfield  {author} {\bibinfo {author} {\bibfnamefont {X.}~\bibnamefont
  {Siemens}}, \bibinfo {author} {\bibfnamefont {V.}~\bibnamefont {Mandic}}, \
  and\ \bibinfo {author} {\bibfnamefont {J.}~\bibnamefont {Creighton}},\ }\href
  {\doibase 10.1103/PhysRevLett.98.111101} {\bibfield  {journal} {\bibinfo
  {journal} {Phys. Rev. Lett.}\ }\textbf {\bibinfo {volume} {98}},\ \bibinfo
  {pages} {111101} (\bibinfo {year} {2007})},\ \Eprint
  {http://arxiv.org/abs/astro-ph/0610920} {arXiv:astro-ph/0610920} \BibitemShut
  {NoStop}%
\bibitem [{\citenamefont {Blanco-Pillado}\ \emph {et~al.}(2018)\citenamefont
  {Blanco-Pillado}, \citenamefont {Olum},\ and\ \citenamefont
  {Siemens}}]{Blanco-Pillado:2017rnf}%
  \BibitemOpen
  \bibfield  {author} {\bibinfo {author} {\bibfnamefont {J.~J.}\ \bibnamefont
  {Blanco-Pillado}}, \bibinfo {author} {\bibfnamefont {K.~D.}\ \bibnamefont
  {Olum}}, \ and\ \bibinfo {author} {\bibfnamefont {X.}~\bibnamefont
  {Siemens}},\ }\href {\doibase 10.1016/j.physletb.2018.01.050} {\bibfield
  {journal} {\bibinfo  {journal} {Phys. Lett. B}\ }\textbf {\bibinfo {volume}
  {778}},\ \bibinfo {pages} {392} (\bibinfo {year} {2018})},\ \Eprint
  {http://arxiv.org/abs/1709.02434} {arXiv:1709.02434 [astro-ph.CO]}
  \BibitemShut {NoStop}%
\bibitem [{\citenamefont {Grishchuk}(1975)}]{Grishchuk1975}%
  \BibitemOpen
  \bibfield  {author} {\bibinfo {author} {\bibfnamefont {L.~P.}\ \bibnamefont
  {Grishchuk}},\ }\href {https://ui.adsabs.harvard.edu/abs/1975JETP...40..409G}
  {\bibfield  {journal} {\bibinfo  {journal} {Soviet Journal of Experimental
  and Theoretical Physics}\ }\textbf {\bibinfo {volume} {40}},\ \bibinfo
  {pages} {409} (\bibinfo {year} {1975})},\ \bibinfo {note} {aDS Bibcode:
  1975JETP...40..409G}\BibitemShut {NoStop}%
\bibitem [{\citenamefont {Lasky}\ \emph {et~al.}(2016)\citenamefont {Lasky}
  \emph {et~al.}}]{Lasky:2015lej}%
  \BibitemOpen
  \bibfield  {author} {\bibinfo {author} {\bibfnamefont {P.~D.}\ \bibnamefont
  {Lasky}} \emph {et~al.},\ }\href {\doibase 10.1103/PhysRevX.6.011035}
  {\bibfield  {journal} {\bibinfo  {journal} {Phys. Rev. X}\ }\textbf {\bibinfo
  {volume} {6}},\ \bibinfo {pages} {011035} (\bibinfo {year} {2016})},\ \Eprint
  {http://arxiv.org/abs/1511.05994} {arXiv:1511.05994 [astro-ph.CO]}
  \BibitemShut {NoStop}%
\bibitem [{\citenamefont {Vagnozzi}(2021)}]{Vagnozzi:2020gtf}%
  \BibitemOpen
  \bibfield  {author} {\bibinfo {author} {\bibfnamefont {S.}~\bibnamefont
  {Vagnozzi}},\ }\href {\doibase 10.1093/mnrasl/slaa203} {\bibfield  {journal}
  {\bibinfo  {journal} {Mon. Not. Roy. Astron. Soc.}\ }\textbf {\bibinfo
  {volume} {502}},\ \bibinfo {pages} {L11} (\bibinfo {year} {2021})},\ \Eprint
  {http://arxiv.org/abs/2009.13432} {arXiv:2009.13432 [astro-ph.CO]}
  \BibitemShut {NoStop}%
\bibitem [{\citenamefont {Arzoumanian}\ \emph {et~al.}(2021)\citenamefont
  {Arzoumanian} \emph {et~al.}}]{NANOGrav:2021flc}%
  \BibitemOpen
  \bibfield  {author} {\bibinfo {author} {\bibfnamefont {Z.}~\bibnamefont
  {Arzoumanian}} \emph {et~al.} (\bibinfo {collaboration} {NANOGrav}),\ }\href
  {\doibase 10.1103/PhysRevLett.127.251302} {\bibfield  {journal} {\bibinfo
  {journal} {Phys. Rev. Lett.}\ }\textbf {\bibinfo {volume} {127}},\ \bibinfo
  {pages} {251302} (\bibinfo {year} {2021})},\ \Eprint
  {http://arxiv.org/abs/2104.13930} {arXiv:2104.13930 [astro-ph.CO]}
  \BibitemShut {NoStop}%
\bibitem [{\citenamefont {Winicour}(1973)}]{Winicour1973}%
  \BibitemOpen
  \bibfield  {author} {\bibinfo {author} {\bibfnamefont {J.}~\bibnamefont
  {Winicour}},\ }\href {\doibase 10.1086/152193} {\bibfield  {journal}
  {\bibinfo  {journal} {The Astrophysical Journal}\ }\textbf {\bibinfo {volume}
  {182}},\ \bibinfo {pages} {919} (\bibinfo {year} {1973})},\ \bibinfo {note}
  {aDS Bibcode: 1973ApJ...182..919W}\BibitemShut {NoStop}%
\bibitem [{\citenamefont {Hogan}(1986)}]{Hogan1986}%
  \BibitemOpen
  \bibfield  {author} {\bibinfo {author} {\bibfnamefont {C.~J.}\ \bibnamefont
  {Hogan}},\ }\href {\doibase 10.1093/mnras/218.4.629} {\bibfield  {journal}
  {\bibinfo  {journal} {Monthly Notices of the Royal Astronomical Society}\
  }\textbf {\bibinfo {volume} {218}},\ \bibinfo {pages} {629} (\bibinfo {year}
  {1986})}\BibitemShut {NoStop}%
\bibitem [{\citenamefont {Deryagin}\ \emph {et~al.}(1986)\citenamefont
  {Deryagin}, \citenamefont {Grigoriev}, \citenamefont {Rubakov},\ and\
  \citenamefont {Sazhin}}]{Deryagin:1986qq}%
  \BibitemOpen
  \bibfield  {author} {\bibinfo {author} {\bibfnamefont {D.~V.}\ \bibnamefont
  {Deryagin}}, \bibinfo {author} {\bibfnamefont {D.~Y.}\ \bibnamefont
  {Grigoriev}}, \bibinfo {author} {\bibfnamefont {V.~A.}\ \bibnamefont
  {Rubakov}}, \ and\ \bibinfo {author} {\bibfnamefont {M.~V.}\ \bibnamefont
  {Sazhin}},\ }\href {\doibase 10.1142/S0217732386000750} {\bibfield  {journal}
  {\bibinfo  {journal} {Mod. Phys. Lett. A}\ }\textbf {\bibinfo {volume} {1}},\
  \bibinfo {pages} {593} (\bibinfo {year} {1986})}\BibitemShut {NoStop}%
\bibitem [{\citenamefont {Caprini}\ \emph {et~al.}(2010)\citenamefont
  {Caprini}, \citenamefont {Durrer},\ and\ \citenamefont
  {Siemens}}]{Caprini:2010xv}%
  \BibitemOpen
  \bibfield  {author} {\bibinfo {author} {\bibfnamefont {C.}~\bibnamefont
  {Caprini}}, \bibinfo {author} {\bibfnamefont {R.}~\bibnamefont {Durrer}}, \
  and\ \bibinfo {author} {\bibfnamefont {X.}~\bibnamefont {Siemens}},\ }\href
  {\doibase 10.1103/PhysRevD.82.063511} {\bibfield  {journal} {\bibinfo
  {journal} {Phys. Rev. D}\ }\textbf {\bibinfo {volume} {82}},\ \bibinfo
  {pages} {063511} (\bibinfo {year} {2010})},\ \Eprint
  {http://arxiv.org/abs/1007.1218} {arXiv:1007.1218 [astro-ph.CO]} \BibitemShut
  {NoStop}%
\bibitem [{\citenamefont {Kobakhidze}\ \emph {et~al.}(2017)\citenamefont
  {Kobakhidze}, \citenamefont {Lagger}, \citenamefont {Manning},\ and\
  \citenamefont {Yue}}]{Kobakhidze:2017mru}%
  \BibitemOpen
  \bibfield  {author} {\bibinfo {author} {\bibfnamefont {A.}~\bibnamefont
  {Kobakhidze}}, \bibinfo {author} {\bibfnamefont {C.}~\bibnamefont {Lagger}},
  \bibinfo {author} {\bibfnamefont {A.}~\bibnamefont {Manning}}, \ and\
  \bibinfo {author} {\bibfnamefont {J.}~\bibnamefont {Yue}},\ }\href {\doibase
  10.1140/epjc/s10052-017-5132-y} {\bibfield  {journal} {\bibinfo  {journal}
  {Eur. Phys. J. C}\ }\textbf {\bibinfo {volume} {77}},\ \bibinfo {pages} {570}
  (\bibinfo {year} {2017})},\ \Eprint {http://arxiv.org/abs/1703.06552}
  {arXiv:1703.06552 [hep-ph]} \BibitemShut {NoStop}%
\bibitem [{\citenamefont {Lee}\ \emph {et~al.}(2021{\natexlab{b}})\citenamefont
  {Lee}, \citenamefont {Mitridate}, \citenamefont {Trickle},\ and\
  \citenamefont {Zurek}}]{Lee:2020wfn}%
  \BibitemOpen
  \bibfield  {author} {\bibinfo {author} {\bibfnamefont {V.~S.~H.}\
  \bibnamefont {Lee}}, \bibinfo {author} {\bibfnamefont {A.}~\bibnamefont
  {Mitridate}}, \bibinfo {author} {\bibfnamefont {T.}~\bibnamefont {Trickle}},
  \ and\ \bibinfo {author} {\bibfnamefont {K.~M.}\ \bibnamefont {Zurek}},\
  }\href {\doibase 10.1007/JHEP06(2021)028} {\bibfield  {journal} {\bibinfo
  {journal} {JHEP}\ }\textbf {\bibinfo {volume} {06}},\ \bibinfo {pages} {028}
  (\bibinfo {year} {2021}{\natexlab{b}})},\ \Eprint
  {http://arxiv.org/abs/2012.09857} {arXiv:2012.09857 [astro-ph.CO]}
  \BibitemShut {NoStop}%
\bibitem [{\citenamefont {Dror}\ \emph {et~al.}(2019)\citenamefont {Dror},
  \citenamefont {Ramani}, \citenamefont {Trickle},\ and\ \citenamefont
  {Zurek}}]{Dror:2019twh}%
  \BibitemOpen
  \bibfield  {author} {\bibinfo {author} {\bibfnamefont {J.~A.}\ \bibnamefont
  {Dror}}, \bibinfo {author} {\bibfnamefont {H.}~\bibnamefont {Ramani}},
  \bibinfo {author} {\bibfnamefont {T.}~\bibnamefont {Trickle}}, \ and\
  \bibinfo {author} {\bibfnamefont {K.~M.}\ \bibnamefont {Zurek}},\ }\href
  {\doibase 10.1103/PhysRevD.100.023003} {\bibfield  {journal} {\bibinfo
  {journal} {Phys. Rev. D}\ }\textbf {\bibinfo {volume} {100}},\ \bibinfo
  {pages} {023003} (\bibinfo {year} {2019})},\ \Eprint
  {http://arxiv.org/abs/1901.04490} {arXiv:1901.04490 [astro-ph.CO]}
  \BibitemShut {NoStop}%
\end{thebibliography}%

\end{document}